# $\mu$LHC: Antimuon Ring and HL-LHC based $\mu^+p$ Collider

D. Akturk[1], A. C. Canbay[2], H. Dagistanli[3], B. Dagli[1], U. Kaya[4], B. Ketenoglu[2], A. Kilic[5], F. Kocak[5], A. Ozturk[6], S. Sultansoy[1,*], I. Tapan[5], F. Zimmermann[7]

[1]*TOBB University of Economics and Technology, Ankara, Turkiye*
[2]*Ankara University, Ankara, Turkiye*
[3]*Private Radikal Educational Institutions, Izmir, Turkiye*
[4]*Ankara University Institue of Accelerator Technologies, Ankara, Turkiye*
[5]*Bursa Uludag University, Bursa, Turkiye*
[6]*Turkish Accelerator and Radiation Laboratory (TARLA), Ankara, Turkiye*
[7]*European Organization for Nuclear Research (CERN), Geneve, Switzerland*

[*]*Corresponding Author:* kayaumt@ankara.edu.tr

## Abstract

Conceptual design and performance evaluation of HL-LHC based antimuon-proton collider ($\mu$LHC) are presented. Leveraging the $\mu$TRISTAN concept based on established J-PARC ultra-cold $\mu^+$ beam technology, $\mu$LHC will give the opportunity to achieve a 5.3 TeV center-of-mass energy, significantly surpassing EIC and LH*e*C. Two booster ring options for $\mu^+$ acceleration, namely, a $\mu$TRISTAN-based and a repurposed LH*e*C ERL-based systems, are explored. Achievable luminosities are predicted to exceed $10^{33}$ cm$^{-2}$s$^{-1}$. The $\mu$LHC offers substantially wider kinematic plane coverage, particularly in small-$x$ and high-$Q^2$ regions, significantly contributing to QCD basics and Higgs boson properties. Its unique potential for BSM physics extends to muon-related phenomena like excited muons, color-octet muons, leptoquarks, and contact interactions. A possible detector concept is also outlined. Given the maturity of ultra-cold $\mu^+$ beam technology, $\mu$LHC is highly feasible for earlier realization than the muon collider, positioning it as a critical tool for the future of high energy physics.

## 1. Introduction

Lepton-hadron collisions have played an exceptional role in the study of the internal structure of matter: proton form factors were first observed in electron scattering experiments [1,2], quarks were first observed at deep inelastic electron scattering experiments of the Stanford Linear Accelerator Center (SLAC) [3,4], European Muon Collaboration (EMC) effect was observed at CERN in deep inelastic muon scattering experiments [5] and so on. Hadron Electron Ring Anlage (HERA) [6], the first and still unique *ep* collider, further explored structure of protons, extended kinematical region by orders, shed light on quantum chromodynamics (QCD) basics and provided parton distribution functions (PDFs) for adequate interpretation of Tevatron and Large Hadron Collider (LHC) data. Concerning Electron Ion Collider (EIC) [7], its *ep* center-of-mass energy is three times lower than HERA (main advantages of EIC are *eA* collisions and polarized beams).

On the other hand, while the electro-weak part of the Standard Model (SM) was completed with the discovery of Higgs boson [8,9], there are still large gaps in the QCD part of the SM: the confinement hypothesis has not been proven from the QCD basics, hadronization and nuclearization phenomena have not been clearly understood. It should be emphasized that Higgs mechanism is responsible for less than 2% of the mass of visible Universe, while remaining is resulting from strong interactions. For these reasons, energy–frontier lepton–hadron colliders are required, since lepton colliders cannot clarify this phenomenon, while hadron colliders have huge background.

Unfortunately, the region of sufficiently small $x$-Bjorken ($< 10^{-4}$) at high $Q^2$ ($> 10$ GeV$^2$), which is very important for understanding QCD basics was not covered by HERA. To investigate this region, lepton–hadron colliders with essentially higher center-of-mass energies are required. Besides, construction of TeV energy lepton–hadron colliders is mandatory to provide PDFs for adequate interpretation of forthcoming data from High Luminosity/High Energy-Large Hadron Collider (HL/HE-LHC) [10,11] and Future Circular Collider/Super proton-proton Collider (FCC/SppC) [12,13] as HERA provided that for Tevatron and LHC. Today, linac–ring type *eh* colliders seem to be the most realistic way to TeV scale in *lh* collisions and Large Hadron-electron Collider (LH*e*C) [14,15] is the most advanced proposal.

Muon-proton colliders were proposed in mid of 1990s [16–18] as an alternative to linac-ring type electron-proton colliders (see reviews [19–26] and references therein). Several years ago, FCC and SppC-based energy frontier muon-hadron colliders have been proposed in [27] and [28], respectively. Recently, several options for LHC/FCC-based $\mu p$ colliders have been considered in [29], $\mu p$ colliders at the BNL have been proposed in [30], HL-LHC and HE-LHC-based $\mu p$ colliders have been considered in [31]. Updated parameters of different $\mu p$ collider proposals are presented in review [32].

As for LHC-based electron-hadron colliders, initially the scheme using Large Electron Positron Collider (LEP) for the electron beam have been considered [33–38]. However, this option was abandoned because of problems related to simultaneous operation of the LHC and LEP in the same tunnel. As a second option, a linear electron accelerator (or collider) tangential to the LHC has been proposed in [39–43]. During last two decades systematic studies on the LHC based *eh* colliders (including accelerator, detector and physics search potential) have been performed by LH*e*C collaboration [44]. Results of the first stage have been published in [14], where three options for *e*-beam were considered: "LEP", 140 GeV one pass linac and 60 GeV energy recovery linac (ERL60). At second stage [15] Energy Recovery Linac (ERL) is considered as the sole version. Recently it was shown that the same performance can be achieved with *e*-ring constructed tangential to the LHC [45,46].

As mentioned above, HL-LHC based multi-TeV scale $\mu p$ colliders were proposed three years ago [31], where construction of muon collider [47,48] tangential to the LHC have been considered. One of the most important problems of muon colliders is cooling and subsequent acceleration of muon beams, construction of cooling demonstrator is scheduled as 2035 in the best (see Figure 12.6.9 in [49]). While a narrow $\mu^-$ beam for muon colliders has not yet been achieved, there is an established technology to create a low-emittance $\mu^+$ beam by using ultra-cold muons [50]. In this context, construction of $\mu^+e^-$ and $\mu^+\mu^+$ colliders at KEK side have been proposed in [51]. Later, the use of $\mu$TRISTAN-like antimuon beam for construction of FCC-*ee* and Circular Electron Positron Collider (CEPC) based $\mu^+e^-$ and FCC-*hh* and SppC based $\mu^+h$ colliders have been considered in [52] and [53], respectively.

Let us mention that the development of ultra-cold anti-muon beams by Japan Proton Accelerator Research Complex (J-PARC) significantly enhances the prospects for muon-hadron colliders. This advancement makes it highly probable that $\mu^+h$ colliders could be established much sooner than $\mu^+\mu^-$ colliders, offering a potential pathway to explore novel physics phenomena at high energies.

This paper proposes the concept of HL-LHC based antimuon-proton ($\mu^+p$) colliders. Section 2 comprehensively addresses the collider's concept and design, encompassing both its accelerator and detector components. Specifically, Subsection 2.1 summarizes the production of ultra-cold positive muons, while their acceleration up to 1 TeV is detailed in Subsection 2.2. The main parameters of $\mu^+p$ colliders are evaluated in Subsection 2.3, and the detector concept is presented in Subsection 2.4. Subsequently, Section 3 provides examples of the collider's physics search potential. This includes the coverage of the kinematic plane ($Q^2 - x$) in Subsection 3.1, a consideration of Higgs production in Subsection 3.2, and an analysis of color-octet muon production in Subsection 3.3. Finally, Section 4 presents the conclusions and recommendations derived from this work.

## 2. Proposed Concept and Design

The proposed concept for a novel $\mu^+p$ collider, called the $\mu$LHC, is primarily inspired by the $\mu$TRISTAN article [51]. This foundational work investigates the feasibility and presents a conceptual design for innovative $\mu^+e^-$ and $\mu^+\mu^+$ colliders at The High Energy Accelerator Research Organization (KEK), based on the ultra-cold antimuon technology developed for the J-PARC muon g-2/EDM experiment. A schematic representation of the conceptual design of the $\mu^+e^-$ and $\mu^+\mu^+$ colliders, as proposed in [51], is depicted in Figure 1.

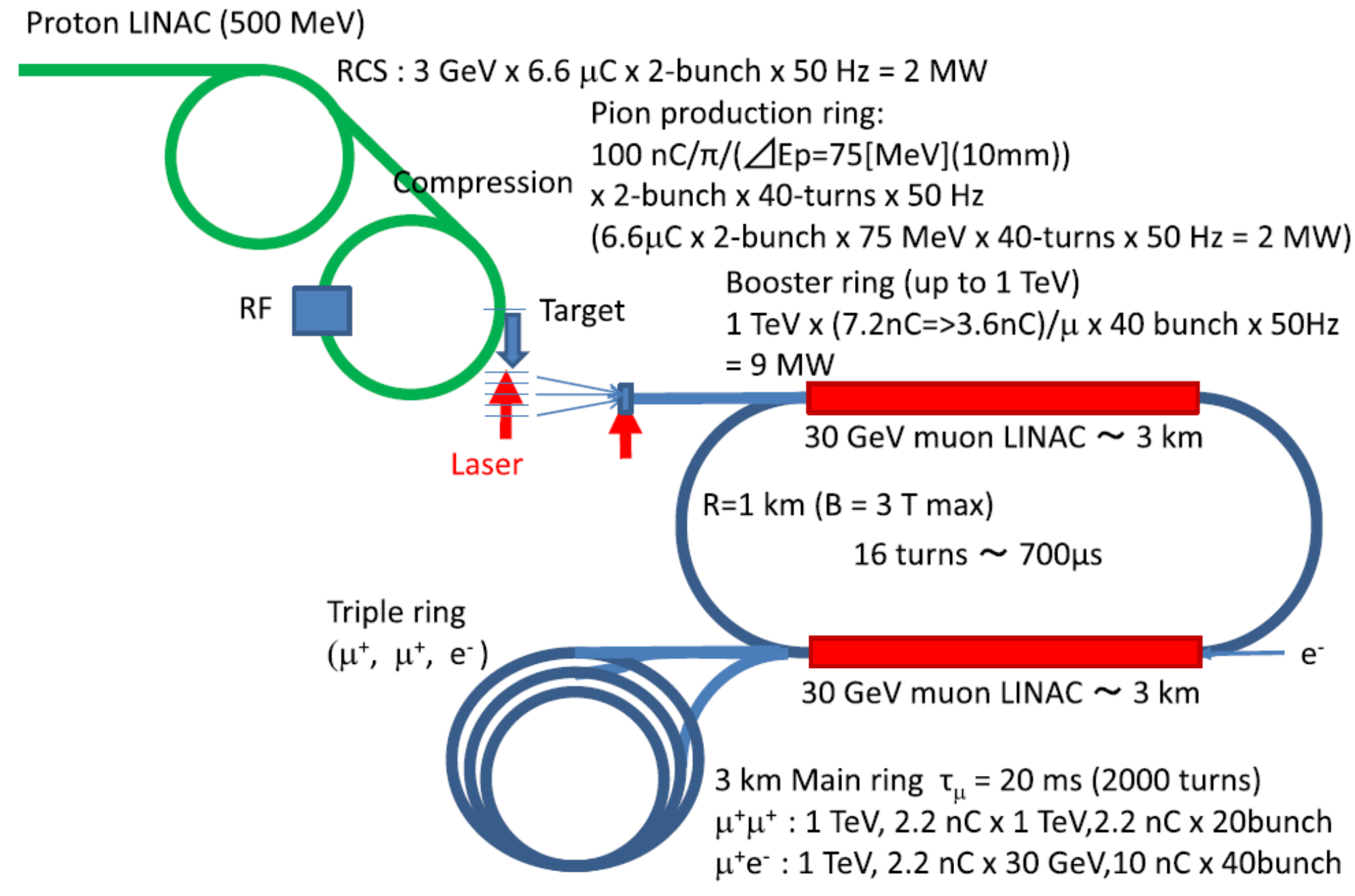

**Figure 1.** Schematic view of conceptual design of $\mu^+e^-$ / $\mu^+\mu^+$ collider (Figure 1 in [51]).

Building upon the antimuon production and acceleration techniques outlined in the original $\mu$TRISTAN design, a compelling muon-proton collider can be conceptualized. This involves tangentially colliding the generated and accelerated antimuons with the existing proton beams within LHC. Such a $\mu$TRISTAN-based $\mu$LHC design would achieve a remarkable center-of-mass energy of 5.3 TeV, resulting from the collision of 1 TeV muons with 7 TeV protons. This energy significantly surpasses the 1.2 TeV center-of-mass energy aimed for by the LH*e*C project, which involves colliding 50 GeV electrons with LHC protons. Consequently, the $\mu$LHC presents an opportunity to construct a lepton-hadron collider capable of providing a center-of-mass energy several times higher than that achievable with the LH*e*C, opening new avenues for high-energy physics research.

### 2.1 Ultra-cold positive muons production

As mentioned in Introduction, one of the primary challenges confronting the realization of muon colliders lies in the efficient production of high-intensity ultra-cold muon beams. While this challenge largely persists for negative muons ($\mu^-$), the production of ultra-cold positive muons (antimuons, $\mu^+$) appears to have found a promising solution.

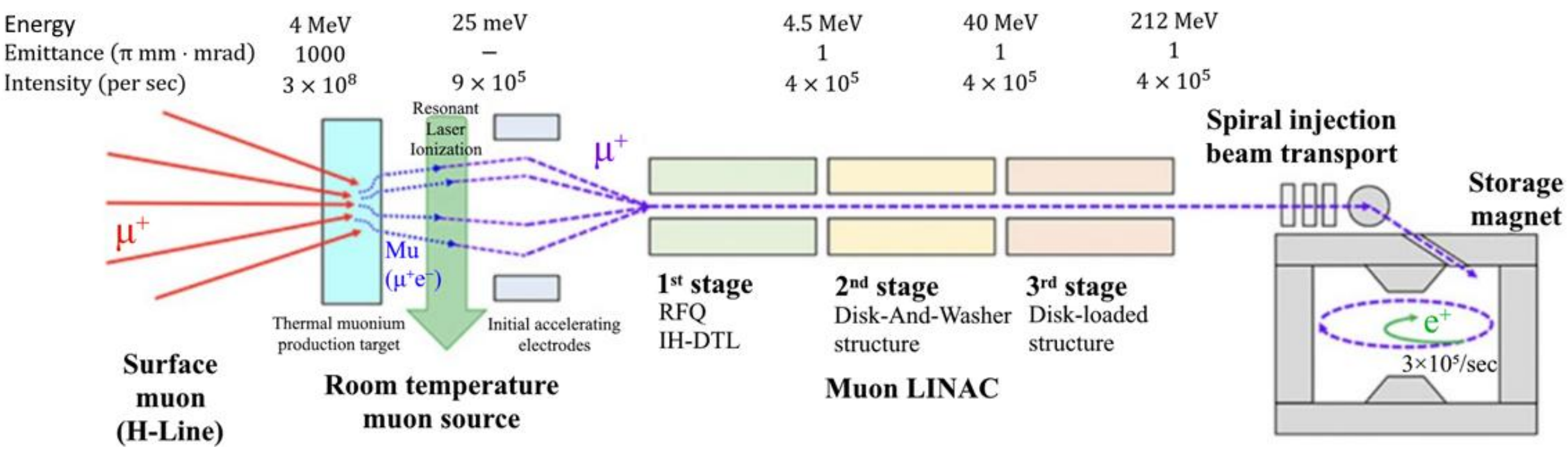

**Figure 2.** Conceptual layout of the J-PARC muon g-2/EDM experiment, highlighting the ultra-cold muon beamline and storage ring [54].

Drawing upon the existing infrastructure at J-PARC's Materials and Life Science Experimental Facility (MLF), the $\mu$TRISTAN project aims to develop a significantly enhanced muon production line. This initiative targets the generation of a much higher muon flux, building upon the ultra-cold muon technology initially developed for the muon g-2 and Electric Dipole Moment (EDM) experiment (as depicted in Figure 2).

According to the design study presented by [51], the ultra-cold muon production rate has been enhanced by a factor of $\mathcal{O}(10^5)$ compared to the J-PARC MLF configuration. This significant improvement results primarily from the efficient use of forward-going pions, which are transported to a large-scale multi-layered stopping target rather than being lost after the initial production.

In contrast to the MLF configuration, where only pions that stopped within a thin region of the graphite target contributed to muon production, the $\mu$TRISTAN design introduces a recirculating proton beam system for pion production. Each proton bunch is recirculated 40 times over a 10 mm graphite target, leading to an enhanced pion yield of 0.016 pions per proton. As depicted in Figure 1, this system integrates a dedicated compression ring with the existing Rapid Cycling Synchrotron (RCS) ring at MLF, enabling protons to repeatedly traverse the target for maximized pion generation. An RF cavity then compensates for the approximately 75 MeV energy loss experienced by protons during each transition within the target, ensuring sustained recirculation.

The produced pions are then directed to a multi-layered tungsten stopping target (Figure 2 in [51]), estimated to consist of ~1000 layers of 1 mm thickness with 1 cm spacing, about 10 meters overall length. This geometry maximizes the stopping efficiency, which is a significant increase over the MLF system. In addition, unlike the MLF system, which utilizes only a small angular acceptance for muon capture, the $\mu$TRISTAN design collects muons over a much broader solid angle.

### 2.2 Boosting ultra-cold muons up to 1 TeV

The booster ring serves as a crucial intermediate acceleration stage, designed to efficiently accelerate positive muons from the muon injection linac to the main ring. In this subsection, two alternative boosting ring options for achieving up to 1 TeV anti-muons will be discussed: the $\mu$TRISTAN-based boosting ring and the LH*e*C's ERL-based boosting ring.

### 2.2.1 $\mu$TRISTAN-based boosting ring

Within the $\mu$TRISTAN collider concept, as depicted in Figure 1, booster ring is conceptualized as a race-track shape that comprises two 3 km-long LINAC sections and arc sections with a radius of 1 km. Positive muons are accelerated from 212 MeV (see Figure 2) to 960 GeV over 16 turns within the ring. However, this acceleration process is inherently challenged by the short lifetime of muons. Despite relativistic time dilation at higher energies, the natural decay of muons leads to an inevitable reduction in beam intensity as the particles circulate and gain energy within the booster ring. According to our analysis, as visually represented in the two-axis graph (Figure 3) showing fraction of survived muons and energy as a function of the number of turns, approximately 82% of the initial muon population remains upon reaching 960 GeV. To achieve 1 TeV instead of 960 GeV, the acceleration for the 3 km long linacs given in the $\mu$TRISTAN design should be 31.25 GeV instead of 30 GeV.

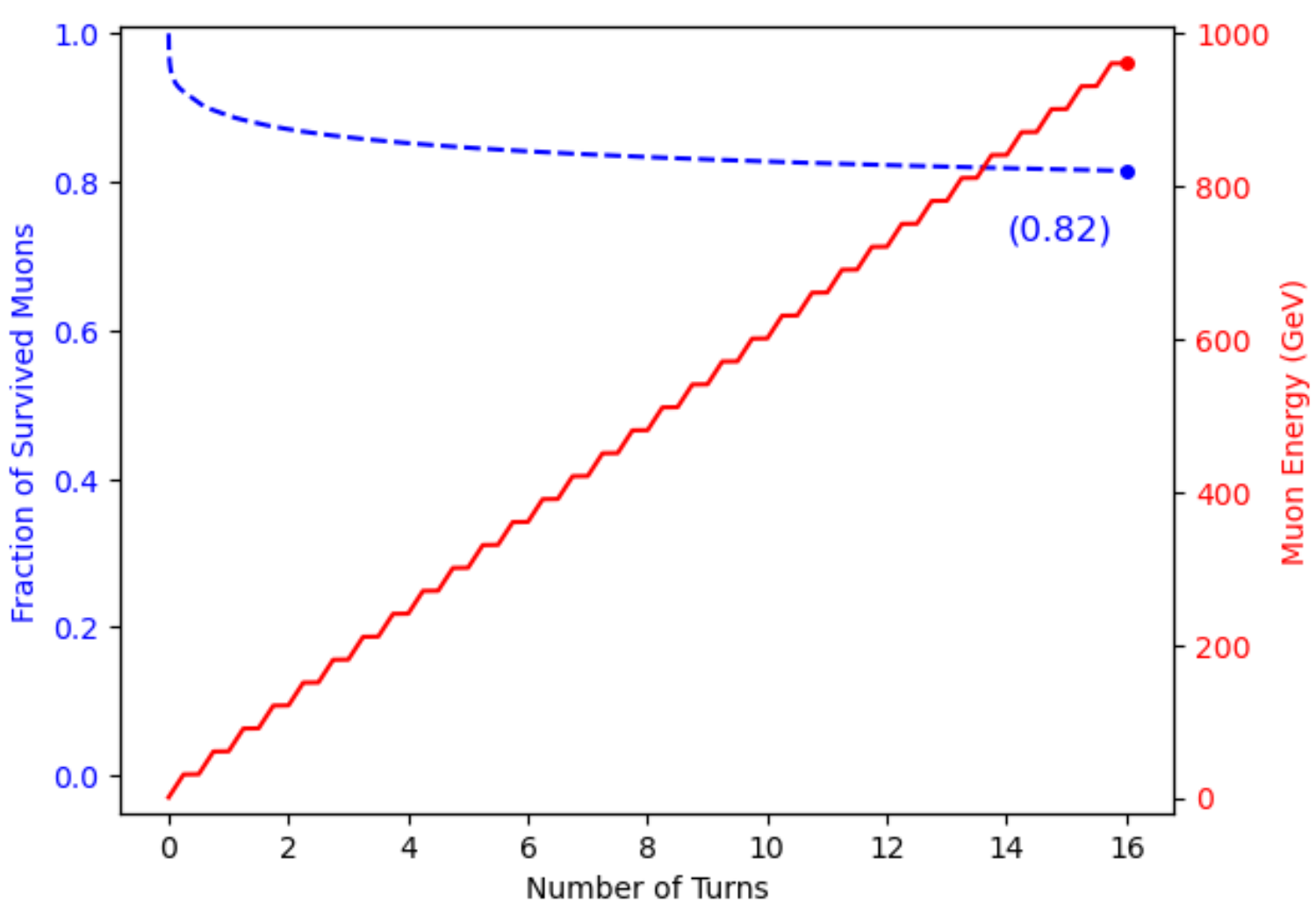

**Figure 3.** Muon beam evolution in the $\mu$TRISTAN booster Ring: energy gain and fraction of survived muons per turn.

While the $\mu$TRISTAN article roughly mentions that about half of the muons survive, our calculation given in Figure 3 indicates a more precise fraction of survived muons, namely, 82%. In the $\mu$TRISTAN study, the muon charge of each bunch injected into the booster ring is stated to be 7.2 nC, which corresponds to $4.5\times10^{10}$ muons. Consequently, the number of surviving muons that are accelerated to 960 GeV is reduced to $3.7\times10^{10}$ muons per bunch. This reduction will directly influence the achievable luminosity of the proposed $\mu^{+}p$ collider.

Following their acceleration in the booster, the positive muons emerge with an energy of 960 GeV and a population of $3.7\times10^{10}$ muons per bunch. In the original $\mu$TRISTAN design, these muons were intended to be injected into two of the three individual 3-km circumference main rings (one for electron beam), which would serve for the execution of $\mu^+e^-$ and $\mu^+\mu^+$ collisions at interaction region surrounded by detectors. However, for the scope of this work, we propose an alternative configuration where this entire system, up to the booster's output, is established at CERN instead of KEK. Our design diverges by suggesting the construction of a single main ring with a circumference of 3 km that would be tangentially aligned with the LHC's proton beam lines. Bird's-eye view of the design is shown in Figure 4.

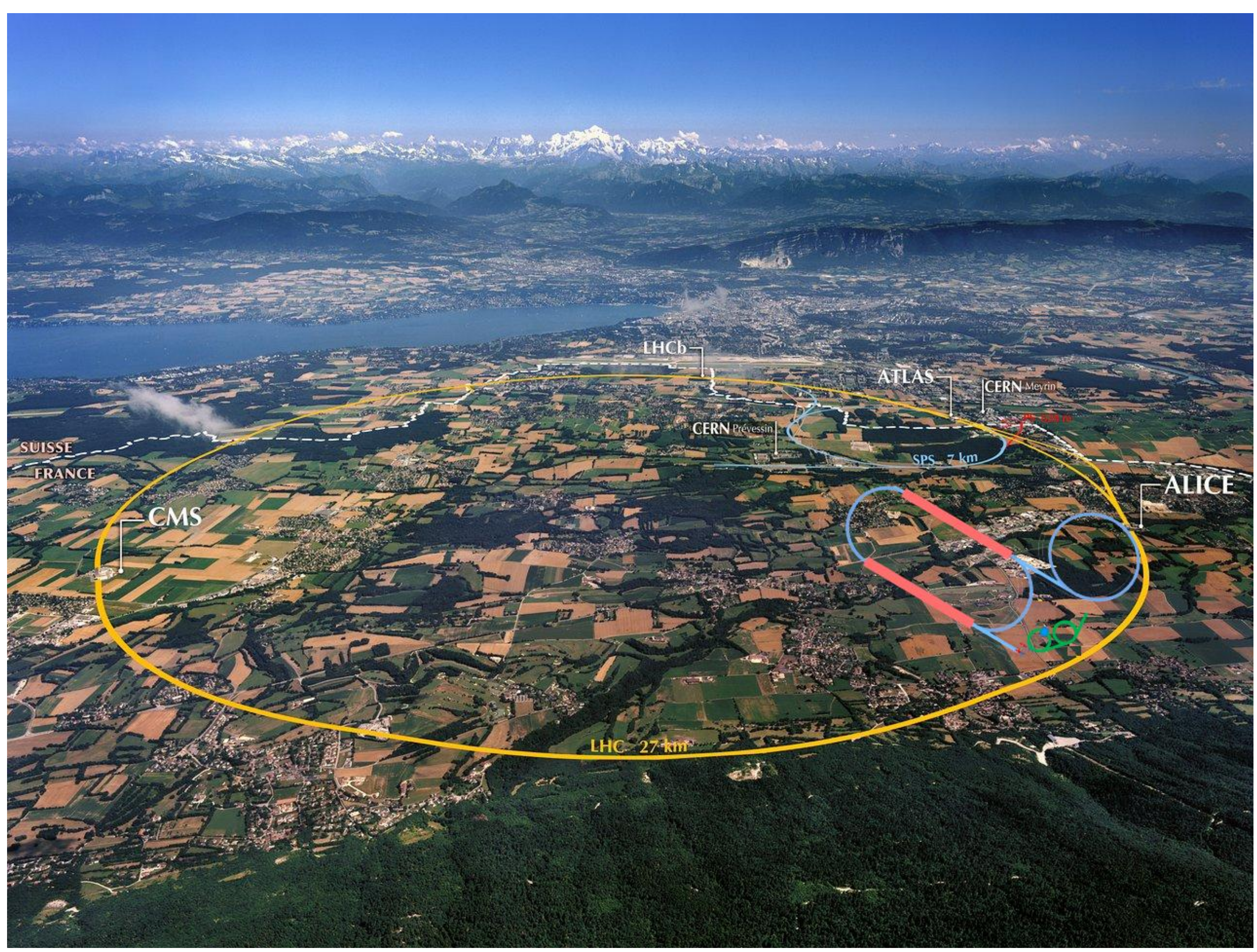

**Figure 4.** Bird's-eye view of the μLHC inspired from $\mu$TRISTAN.

### 2.2.2 LH*e*C's ERL-based boosting ring

The LH*e*C project [15] proposes an electron-hadron collider at CERN, designed to operate in conjunction with the HL-LHC. By colliding high-energy electrons with the proton or heavy-ion beams of the LHC at the TeV scale, the LH*e*C aims to probe fundamental particle interactions and matter structure. The LH*e*C project's electron acceleration facility is designed as a multi-turn ERL system (as visualized in Figure 5). The baseline design, as detailed in [15], envisions a "race-track" layout utilizing two main linacs, each approximately 1 km long, which are connected by three recirculation arcs to facilitate the multi-turn operation. These linacs are designed to provide an acceleration of 10 GeV per pass. To achieve the desired final energy, electrons are typically recirculated three times after injection.

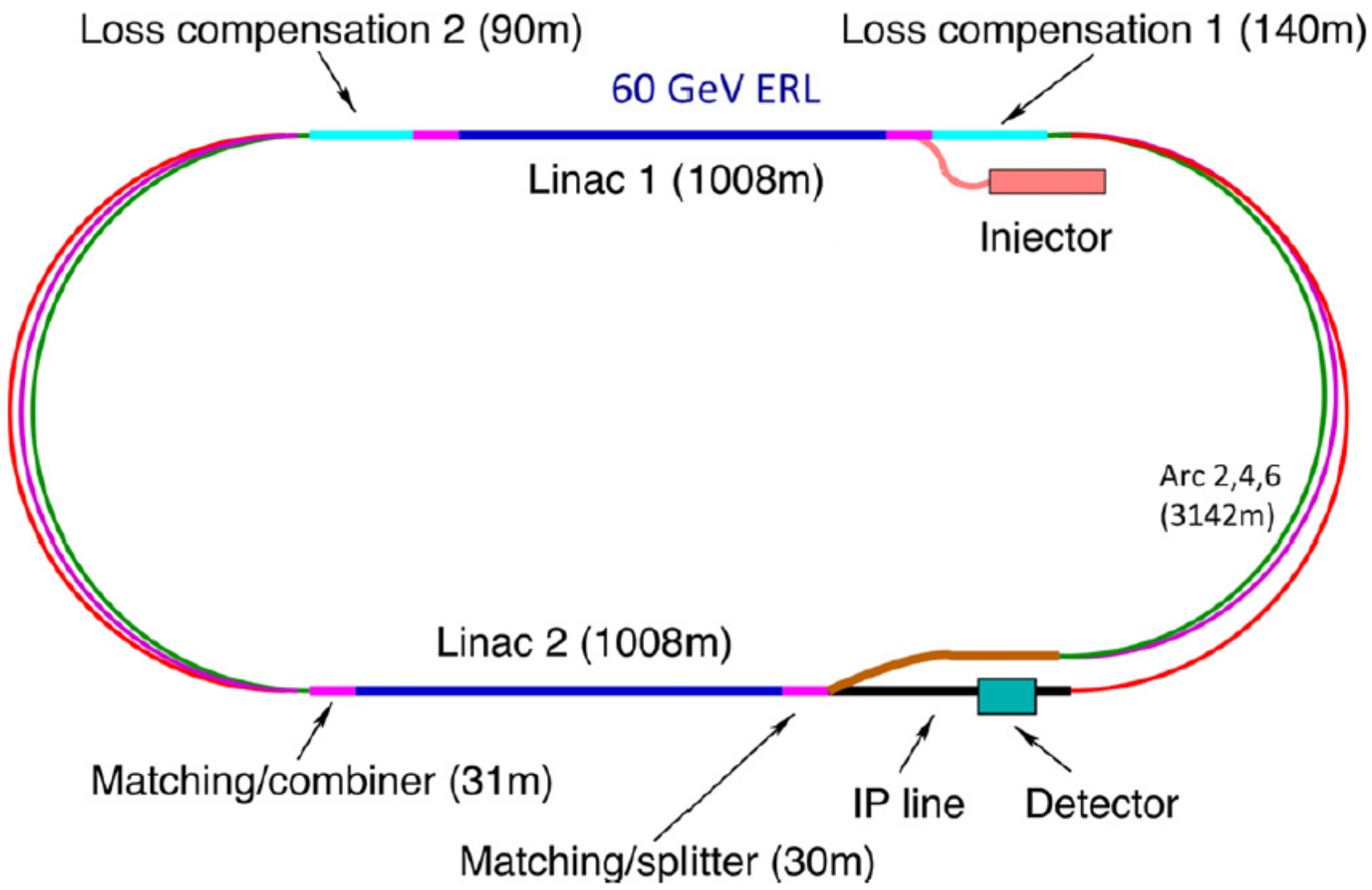

**Figure 5.** Schematic view of the three-turn LHeC configuration at 60 GeV final energy scale with two oppositely positioned electron linacs and three arcs housed in the same tunnel.

Maintaining beam energy requires continuous RF power to compensate synchrotron radiation losses, which intensify significantly with increasing electron energies. Such considerable power consumption thus becomes a critical design parameter, directly influencing the operational limits of the LH*e*C. The LH*e*C project initially considered a 60 GeV electron beam energy as its default [14]. However, the 50 GeV option is selected as the new default [15], primarily driven by a strict 100 MW wall-plug power constraint.

As seen in Figure 1 and Figure 5, ERL tunnel to be built for LH*e*C has similar structure of the booster ring of the $\mu$TRISTAN project, which plans to accelerate anti-muons up to 1 TeV energies. While both facilities share a similar "race-track" shape, a key distinction lies in the LH*e*C incorporating two parallel linacs, each approximately 1 km in length instead of 3 km. Therefore, the LH*e*C's ERL tunnel can be repurposed as a muon booster ring for an ERL-based $\mu$LHC. A schematic view of this proposal is given in Figure 6.

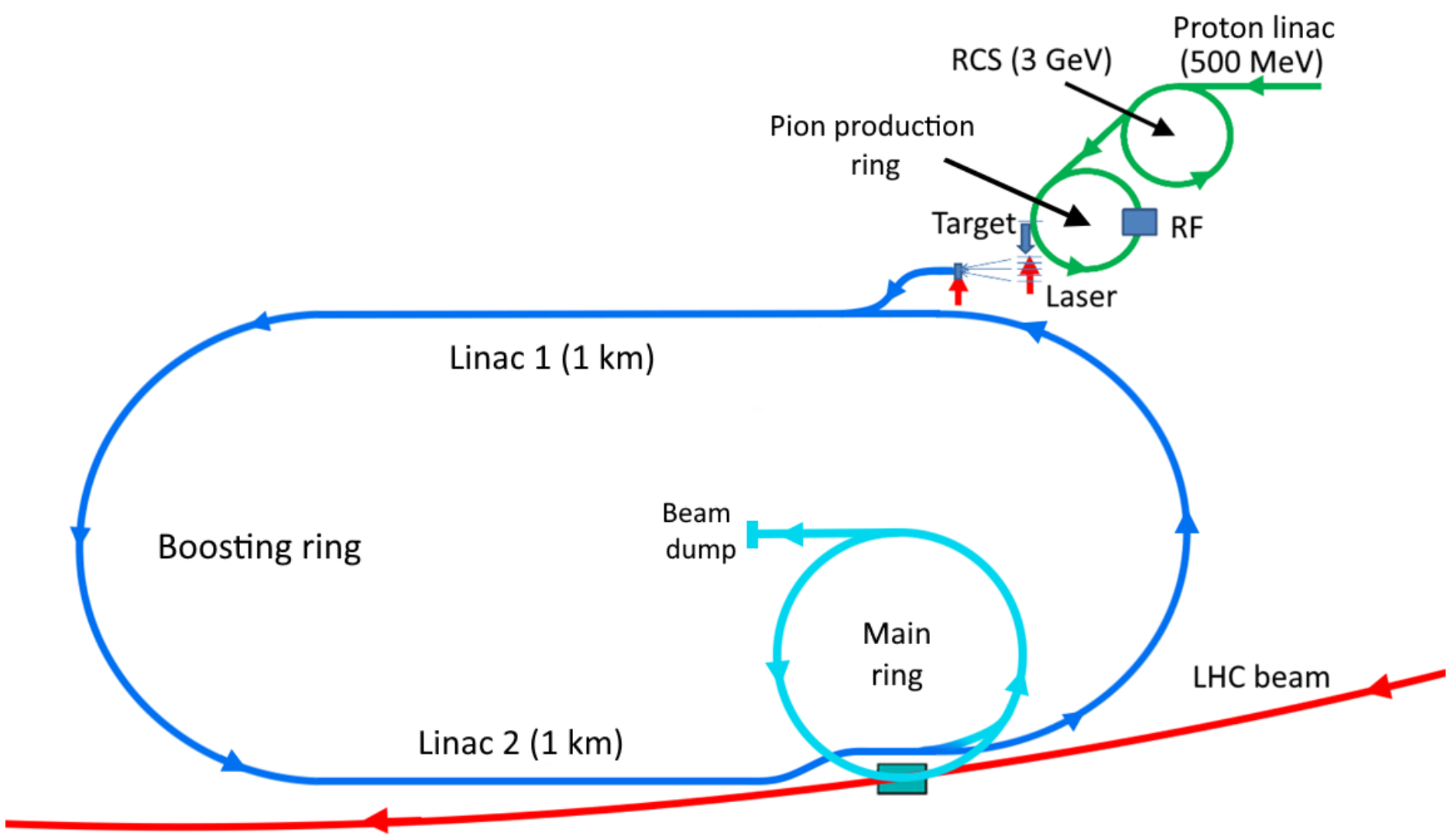

**Figure 6.** Schematic view of the $\mu$LHC, showing a booster ring and main ring compactly integrated within the ERL tunnel, in a manner similar to that proposed for $\mu$TRISTAN.

In a recent study [55], a simplified "phase-one" LH*e*C have been proposed. This new proposal suggests a 20 GeV, single-pass ERL with a higher beam current of 60 mA and a circumference one-third that of the LHC. This design choice is motivated by the existing availability of single-pass ERL technology and the significant simplification it offers, particularly in the machine-detector interface due to reduced synchrotron radiation. Furthermore, this "phase-one" LH*e*C configuration is noted to have substantially lower building and operational costs compared to the final LH*e*C design. One pass option would allow *lh* collision experiments with electrons accelerated up to 20 GeV to commence significantly earlier. Subsequently, as a next phase, either a 50 GeV multi-pass linac option could be constructed, or booster ring of $\mu$LHC could be directly installed within the ERL tunnel.

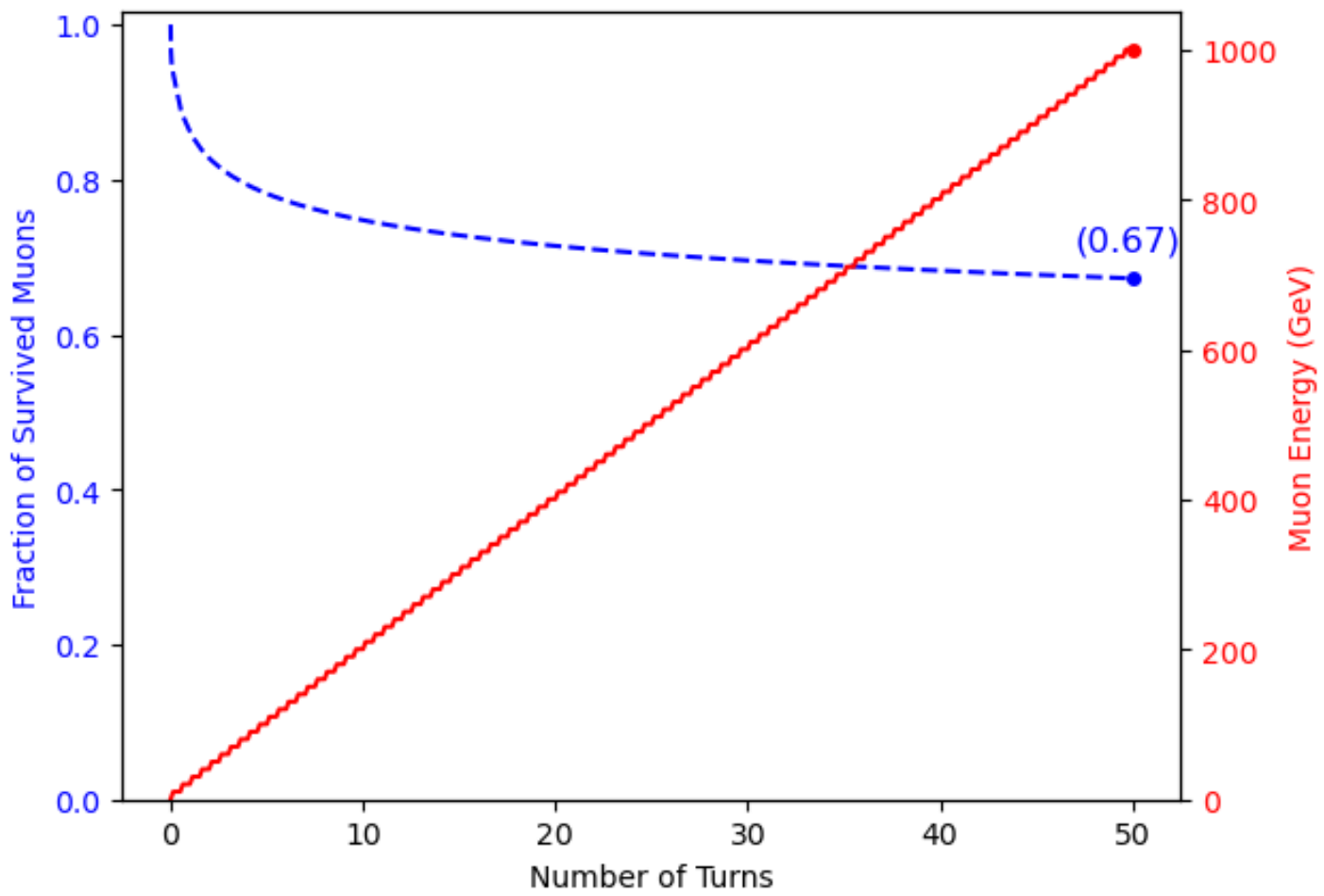

**Figure 7.** Muon beam evolution in the ERL-based booster Ring: energy gain and fraction of survived muons per turn.

Differing from the previous case, acceleration of positive muons from 212 MeV to 1 TeV requires 50 turns within the ERL based booster ring. Fraction of survived muons and energy as a function of the number of turns are shown in Figure 7. It is seen that approximately 67% of the initial muon population remains upon reaching 1 TeV. Consequently, the number of surviving muons that are accelerated to 1 TeV is reduced from $4.5\times10^{10}$ to $3.0\times10^{10}$ muons per bunch.

### 2.2.3 Magnet Lattice and Acceleration Scheme

High-energy muon beam has two main challenges: efficient ultra-cold muon production and rapid acceleration within the muon's short lifetime [56, 57]. The $\mu$TRISTAN concept [51] provides a viable solution for the "production" stage by using ultra-cold anti-muon technology developed at J-PARC. Therefore, the primary challenge for the proposed $\mu$LHC remains the "rapid acceleration" phase.

The original $\mu$TRISTAN proposal suggests a booster ring to accelerate muons from hundreds of MeV to 1 TeV energies over 16 turns. This design features a race-track geometry with 3 km linac sections and 1 km radius arcs, employing "alternative gradient bending magnets of 3 T" [51]. While the specific lattice type is not defined in detail, the requirement to handle large momentum spreads points toward a Fixed-Field Alternating Gradient (FFAG) design.

For acceleration technology, the Muon Collider baseline currently favors a Hybrid Rapid Cycling Synchrotron (Hybrid-RCS) [58]. This approach interleaves fixed-field superconducting magnets with fast-ramping normal conducting ones to accelerate a single bunch [56]. However, the $\mu$LHC scheme presents a different set of constraints. The $\mu$TRISTAN production scheme inherently generates a train of 40 bunches [51]. While a bunch merging system could theoretically reduce this train to a single bunch to enable RCS compatibility, avoiding such complex manipulation requires an accelerator capable of handling the multi-bunch structure. Without merging, ramping magnets for a continuous 40-bunch train would place extra challenging demands on power supplies and ramp rates.

Given these constraints, the vertical excursion Fixed-Field Alternating Gradient (vFFA) accelerator [59], also noted as an alternative in recent MuCol studies [58], is a more appropriate choice for the $\mu$LHC booster. Since the vFFA uses static magnetic fields with the orbit shifting vertically with energy, it allows for isochronous acceleration without the limitations of magnet ramping.

Regarding the LHeC's ERL-based $\mu$LHC boosting ring option, it is evident that the standard bending arcs designed for 20 GeV and 50 GeV electron recirculation are unsuitable for the $\mu$LHC. To implement vFFA within the tunnel constraints and to mitigate the cost of large-aperture magnets, a cascaded arc scheme could be considered, utilizing the tunnel's capacity for multiple arcs. In such a scenario, separate vFFA arcs with optimized apertures would cover different energy ranges (e.g., low, mid, and high energy). Future technical design studies should investigate the feasibility of matching this lattice with the proposed RF systems of LHeC's ERL.

## 2.3 Main Parameters of $\mu^+p$ Colliders

Upon injection into the main ring, the muon beam, having been accelerated to up to 1 TeV by the boosting ring, exhibits specific characteristics crucial for high-luminosity $\mu p$ collisions. A beam composed of 40 bunches [51] of ultra-cold muons then enters the main ring. Considering the 3 km circumference of the main ring, the bunch-to-bunch distance would be determined to be 75 meters, corresponding to a bunch spacing of 250 ns, which can be synchronized with one in ten of the 25 ns of bunch spacing of proton beams at the LHC.

Concerning the lifetime of muons within the main ring, the relativistic lifetime of muons injected into the main ring at ~1 TeV is approximately 20 ms. When these highly relativistic muons circulate in a main ring with a circumference of 3 km, each turn takes approximately 10 μs. Consequently, a muon can complete about 2000 turns within the ring during its lifetime. Furthermore, considering the exponential decay inherent in the muon's nature, the average number of muons present in the ring throughout this 2000-turn circulation period is determined to be approximately 0.63 times the initial injected number of muons.

In the case of $\mu$TRISTAN-based boosting ring, given an injection $3.7\times10^{10}$ muons per bunch into the main ring, the average muon population in the main ring is approximately $2.3\times10^{10}$ muons per bunch during its lifetime. Similarly, for the LH*e*C's ERL-based boosting ring option, the muon population in the bunch is observed to decrease from an initial $3.0\times10^{10}$ muons to an average of $1.9\times10^{10}$ muons. However, considering the initial muon number utilized in this work, the precise calculation of the muon population reduction within the main ring was assigned to the A luminosity optimiser for High Energy Physics (AloHEP) software.

AloHEP [60–62] is a dedicated software tool for accelerator design, focusing on beam dynamics calculations on interaction region. It computes crucial parameters like luminosity, tune-shift, and disruption for various collider types (linear, circular, linac-ring) and diverse beams (electrons, positrons, muons, protons, nuclei). Utilizing accelerator design parameters, AloHEP simulates interaction region to calculate luminosity. It accounts for significant luminosity-altering effects such as the pinch effect, particle decay, hourglass effect etc.

The parameters of antimuon beam are presented in Table 1. The numbers of antimuons are taken from the muon population exiting the boosting ring, discussed in the previous subsection. The remaining beam parameters are adapted from the $\mu^+$ beam detailed in the $\mu^+e^-$ collider proposal of [51].

**Table 1.** Main parameters of $\mu^+$ beam.

| Parameter | μTRISTAN-based | ERL-based |
|---|---|---|
| Number of muons per bunch [$10^{10}$] | 3.7 | 3.0 |
| Beam energy [GeV] | 1000 | |
| $\beta_x$ @ IP [cm] | 3 | |
| $\beta_y$ @ IP [cm] | 0.7 | |
| Bunch length [mm] | 2 | |
| Norm. horizontal emittance [μm] | 4 | |
| Norm. vertical emittance [μm] | 4 | |
| Number of bunches per ring | 40 | |
| Collision frequency [MHz] | 4 | |
| Circumference [km] | 3 | |

The proton beam parameters utilized for luminosity calculations in the $\mu$LHC collider are adopted from the ERL60-upgraded proton parameters developed for the LH*e*C and are presented in Table 2 (Table 2.11 in [12]).

**Table 2.** ERL60-upgraded proton parameters.

| Parameter | p |
|---|---|
| Number of Particle per Bunch [$10^{10}$] | 22 |
| Beam Energy [TeV] | 7 |
| β function @ IP [cm] | 7 |
| Norm. Emittance [μm] | 2 |
| Number of Bunches per Ring | 2760 |
| Rev. frequency [Hz] | 11245 |
| Circumference [km] | 26.7 |

Using the antimuon and proton parameters given above, luminosity values for two options were calculated within the AloHEP software, with beam size matched accordingly. The resulting main parameters of $\mu^+p$ collider for both cases are presented in Table 3.

**Table 3.** Main parameters of the $\mu$TRISTAN- and ERL-based $\mu^+p$ collider.

| Parameter | $\mu$TRISTAN-based | | ERL-based | |
|---|---|---|---|---|
| √s [TeV] | 5.3 | | 5.3 | |
| L [$10^{33}$ cm$^{-2}$s$^{-1}$] | 8.8 | | 7.2 | |
| Parameter | $\mu^+$ | p | $\mu^+$ | p |
| $\sigma_{x,y}$ [μm] | 4.3 | 4.3 | 4.3 | 4.3 |
| $\xi_{x,y}$ [$10^{-4}$] | 600 | 22 | 600 | 18 |

## 2.4 The Detector Concept

A proposed muon and hadron colliders for both the SM (especially for clarifying QCD basics) and Beyond Standard Model (BSM) phenomena is based on colliding a 7 TeV proton beam against a 1 TeV muon beam. The $\mu h$ detector will be constructed for the detection of the producing particles from this collision. The detector is configured superconducting solenoid magnet (SSM) for precise momentum measurements for charged tracks. Figure 8 shows the preliminary design of a general purpose $\mu h$ detector structure which is similar to the previously proposed by Acosta et al. [57]. The detector given here is the classical cylindrical layout typical for multipurpose detectors of asymmetric collisions. The scattered parton (e.g., jets) and muon are mostly going toward the backward direction [63]. Therefore, asymmetric detector structure consists of a tracker (SiT) to provide precision timing and position information, a particle identification (PID) system located outside of the silicon tracker, calorimeters (ECAL and HCAL) to precisely measure the energies of the particles and muon spectrometer (MUON) to measure muons with proper resolution. These sub-detectors surround the beam pipe. On the beam pipe outside the detector, Roman Pots (RPs) in the far forward direction and a muon system (MS) in the far backward direction are mounted to detect the scattered protons and muons respectively.

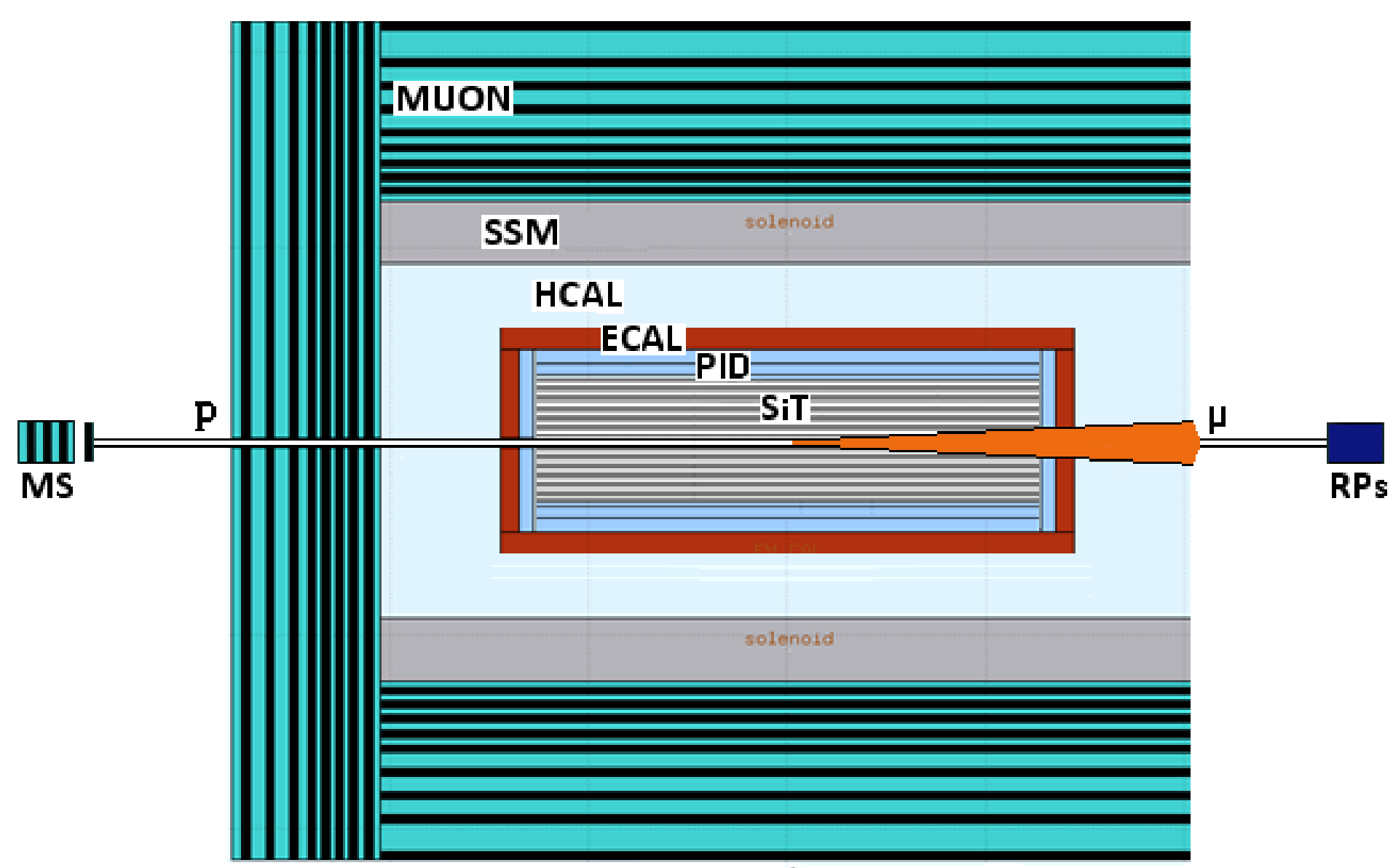

**Figure 8.** Schematic view of the designed detector at muon-proton colliders.

Beam-induced backgrounds (BIB) are one of the main challenges in the design of the $\mu^{+}p$ detector. Decays products of the muon beams cause the presence of a large amount of the background particles (such as photons, electrons and neutrons) in the detectors. It is necessary to distinguish the generated particles result of the muon-proton collisions from an intense BIB coming from the muon decay products. For this reason, a shielding nozzle is used to protect the innermost detector elements. The shielding tungsten nozzle cladded with borated polyethylene is placed to the muon coming side, as seen as brown conical shape in Figure 8.

However, the nozzle affects the physics performance by limiting the angular acceptance of the detector. In addition to this, particles creating from energy loss by multiple scattering of outgoing high-energy muons passing through the shielding cone will disturb measurements.

The GEANT4 and FLUKA simulation codes have been used to simulate the energy loss of high energy muons passing through the 6-meter-long tungsten shielding cone. Figure 9 shows the simulated energy loss of 1 TeV muons traversing such a tungsten cone.

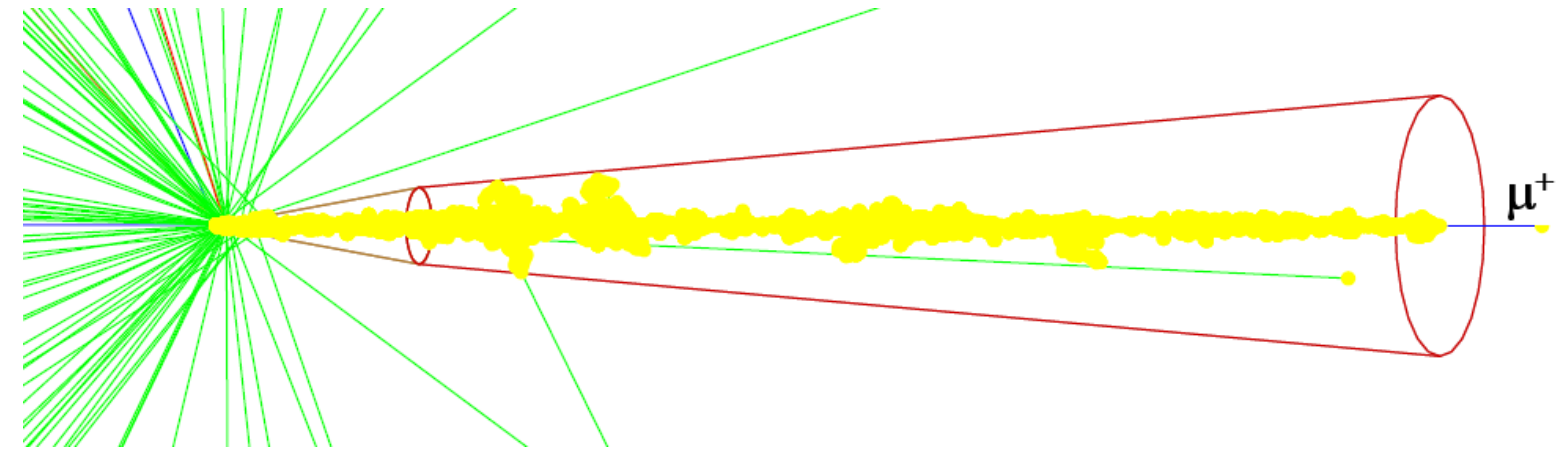

**Figure 9.** Geant4 simulation of energy deposition in the tungsten cone.

The simulations give the average deposited and passing muons energies around as 19 % and 81 % respectively through the tungsten shielding cone volume, as seen in Figure 10.

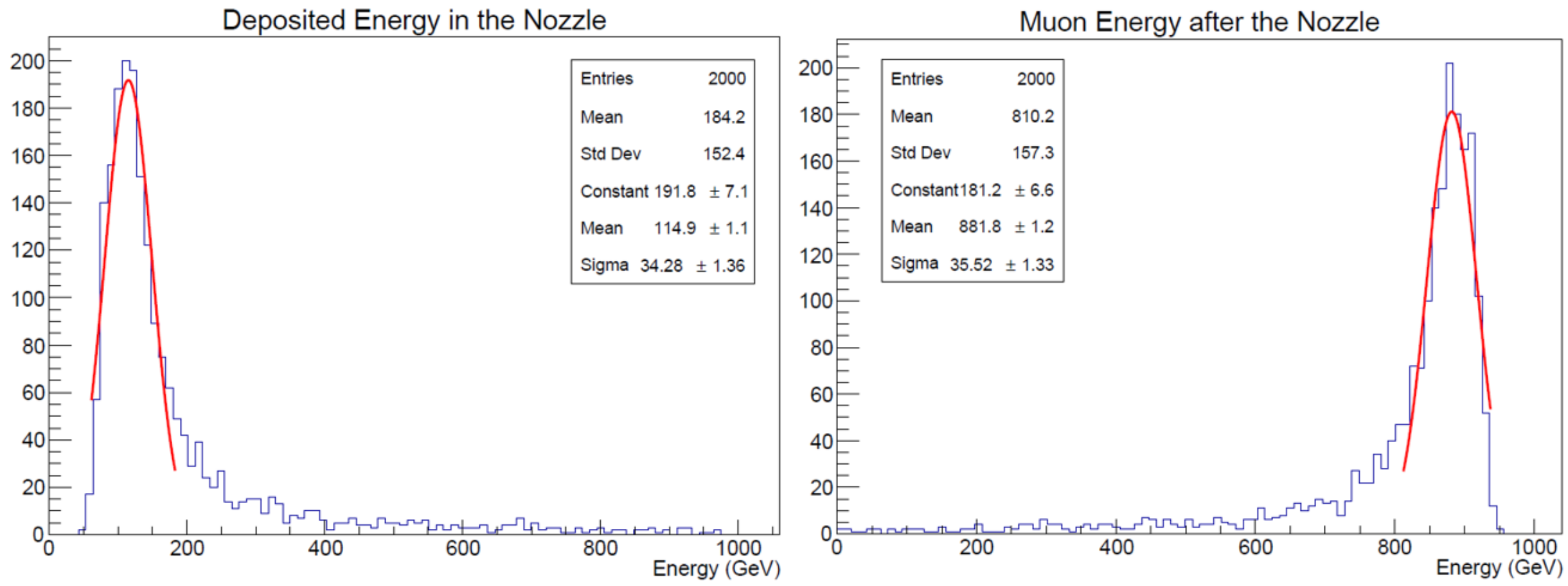

**Figure 10.** Energy distributions of 1000 GeV muons through the nozzle.

Shielding studies will be detailed to optimize the overall physical performance of the detector. The variations of the energy depositions at the different muon energies traversing such a cone made of different stopping target materials will be studied.

Systematic studies of the detector performance for the LHC/FCC/SppC based muon-proton colliders are necessary for long-term planning of High Energy Physics. The realistic dimensions of the detector and its sub-components could be determined using detailed GEANT4 and FLUKA simulations to get the detector's response to particles generated from the muon-proton collisions.

## 3. Physics Potential of $\mu^+p$ colliders

Multi-TeV center-of-mass energy μh colliders have a huge potential for both the SM (especially QCD part) and BSM physics searches (see review articles [29,57,64] and references therein). In following subsections, small $x$-Bjorken region, Higgs boson and color-octet muon productions are considered. Main parameters of *lh* colliders used in this subsection are presented in Table 4. Parameters of HERA and EIC are taken from Particle Data Group (PDG) [65], parameters of ERL20 option of LH*e*C is taken from [55], for ERL50 option of LH*e*C latest parameters presented in [66] are used. Parameters of $\mu$LHC are evaluated in previous subsection.

**Table 4.** Center-of-mass energies and luminosities of HERA, EIC, LH*e*C and $\mu$LHC.

| Collider | $\sqrt{s}\,[TeV]$ | $L[cm^{-2}s^{-1}]$ | $L_{int}[fb^{-1}]$ |
|---|---|---|---|
| HERA | 0.319 | $7.5 \times 10^{31}$ | 0.8 (total) |
| EIC | 0.105 | $1.0 \times 10^{34}$ | 100/year |
| LHeC (ERL 20) | 0.748 | $6.0 \times 10^{33}$ | 60/year |
| LHeC (ERL 50) | 1.18 | $1.4 \times 10^{33}$ | 14/year |
| μLHC ($\mu$TRISTAN based) | 5.3 | $8.8 \times 10^{33}$ | 88/year |
| μLHC (ERL based) | 5.3 | $7.2 \times 10^{33}$ | 72/year |

### 3.1 Small *x*-Bjorken

Investigation of the region of sufficiently small *x*-Bjorken (< $10^{-4}$) at high $Q^2$ ($> 10\ GeV^2$) is crucial for understanding QCD basics. For lepton-hadron colliders, the relation between *x*-Bjorken and $Q^2$ is given as $Q^2 = 4x_B E_\mu E_p$. Achievable *x*-Bjorken values at $Q^2 = 10\ GeV^2$ and $Q^2$ values at $x = 10^{-4}$ for HERA, EIC, LH*e*C and $\mu$LHC are presented in Table 5. As can be seen from the table, the proposed colliders will allow detailed exploration of the relevant ($x - Q^2$) region.

**Table 5.** Achievable $x - Bj\ddot{o}rken$ values at $Q^2 = 10\ GeV^2$ and $Q^2$ values at $x = 10^{-4}$.

| Collider | $\sqrt{s}\,[TeV]$ | $x - Bjorken$ | $Q^2\ [GeV^2]$ |
|---|---|---|---|
| HERA | 0.319 | $1.0 \times 10^{-4}$ | 10 |
| EIC | 0.145 | $4.8 \times 10^{-4}$ | 2.1 |
| LHeC (ERL 20) | 0.748 | $1.8 \times 10^{-5}$ | 56 |
| LHeC (ERL 50) | 1.18 | $7.2 \times 10^{-6}$ | 140 |
| μLHC | 5.3 | $3.6 \times 10^{-7}$ | 2800 |

Coverage of the kinematic plane in deep inelastic lepton–proton scattering by some initial fixed-target experiments using electrons (SLAC) and muons (NMS and BCDMS experiments), and by the *ep* colliders, the EIC (green), the HERA (yellow), the LH*e*C (blue) and the FCC-*eh* (brown), are presented Figure 11 (Figure 1 in [15]). Here, we added $\mu$LHC collider coverage. It is seen that $\mu$LHC covers essentially wider range of small-*x* and high-*Q*$^2$ than the LH*e*C and even the FCC-*eh*.

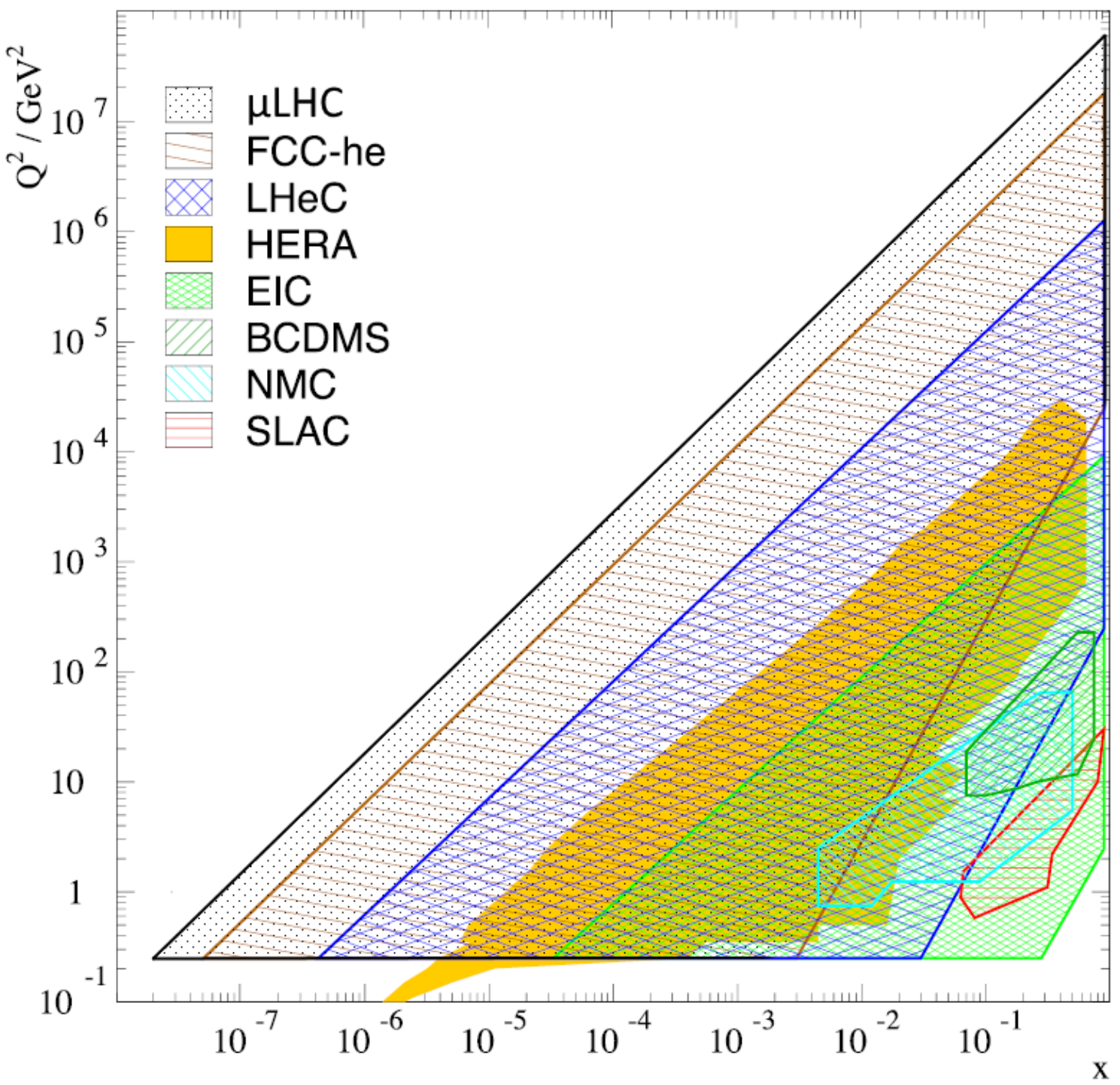

**Figure 11.** Coverage of the kinematic plane in deep inelastic lepton–proton scattering.

### 3.2 Higgs production

Cross-sections for Higgs boson production in $\mu^{+}p$ collisions via W and Z fusion processes are presented in Figure 12. The numbers of Higgs bosons produced per working year ($10^{7}$ s) for colliders under consideration are given in Table 6. It is clearly seen that $\mu$LHC colliders are much more advantageous than the LH*e*C.

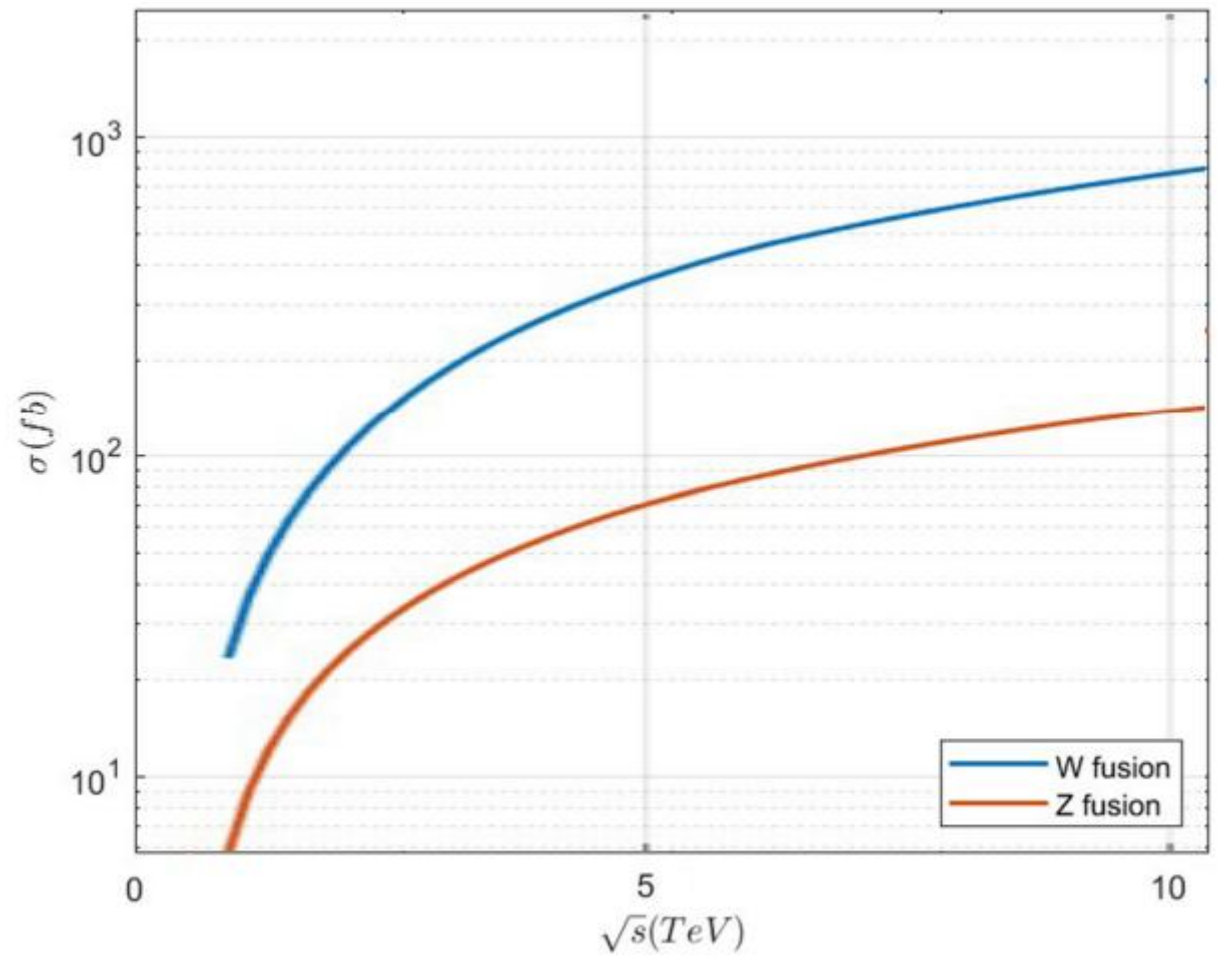

**Figure 12.** Cross sections for Higgs boson production in $\mu^{+}p$ collisions via W and Z fusion processes.

**Table 6.** Number of Higgs bosons produced per working year.

| Collider | $\sqrt{s}\ [TeV]$ | $L_{int}[fb^{-1}]$ | $N_{W fusion}$ | $N_{Z fusion}$ |
|---|---|---|---|---|
| LHeC (ERL 20) | 0.748 | 60 | 1260 | 200 |
| LHeC (ERL 50) | 1.18 | 14 | 1020 | 200 |
| μLHC (μTRISTAN based) | 5.3 | 88 | 33900 | 6600 |
| μLHC (ERL based) | 5.3 | 72 | 27500 | 5500 |

### 3.3 Search for Color-Octet Muons at *μ*LHC

The lepton (and quark) composite paradigm, in contrast to the SM, offers a highly compelling theoretical framework. It posits that leptons may not be fundamental particles, but rather, they may be composed of more fundamental constituents, referred to as "*preons*" (see [67] and reference therein). In certain composite frameworks, each SM lepton (which is a color-singlet under $SU(3)_c$) possesses a strongly interacting color-octet partner. This partner is frequently referred to as a "leptogluon". The color-octet muon, *μ8*, is a constituent of this structure, bearing a color charge in the octet (adjoint) representation of QCD, while concurrently sharing intrinsic electromagnetic and weak quantum numbers with the SM muon [68]. This distinctive hybrid nature of *μ8* renders it an optimal candidate for examining composite structure of leptons. The interaction Lagrangian of leptogluons ($l_8$) with leptons and gluons can be written as [69]:

$$\mathcal{L} = \frac{1}{2\Lambda}\left\{\overline{l_8^{\alpha}}\, g_S G_{\mu\nu}^{\alpha} \sigma^{\mu\nu}(\eta_L l_L + \eta_R l_R) + h.c.\right\}$$

Here, $g_S$ is the strong coupling constant, $\Lambda$ is the compositeness scale, $G_{\mu\nu}$ is the gluon field strength tensor, $l_{L/R}$ is the left/right spinor components of the lepton, $l$ is SM lepton, $\sigma^{\mu\nu}$ is the antisymmetric tensor and $\eta_{L/R}$ is the chirality factor.

In the context of proton colliders, such as the LHC and the FCC, the production of color-octet muons is typically constrained to pair production [69]. However, in *μp* colliders, the direct interaction between the muon and gluon inside the proton allows for the resonance production as $\mu\, g\ \rightarrow\ \mu 8$. It is evident that this production channel exhibits a remarkably elevated cross-section, a phenomenon that can be attributed to its notable high-energy gluon ratio and intricate resonance structure.

In this subsection, the search for *μ8* at the *μ*LHC collider is investigated. The execution of event productions was accomplished through the utilization of MadGraph5 [70], employing the model file [71] for leptogluons that was prepared in FeynRules [72]. The compositeness scale, $\Lambda = m_{\mu 8}$, was employed to facilitate the process ($E_\mu = 1$ TeV and $E_p = 7$ TeV). Figure 13 presents representative Feynman diagrams for single and pair production, as well as plots of cross-sections versus *μ8* mass for both single production at *μ*LHC and pair production at HL-LHC. A salient observation is the conspicuously elevated cross sections for single production in comparison to pair production.

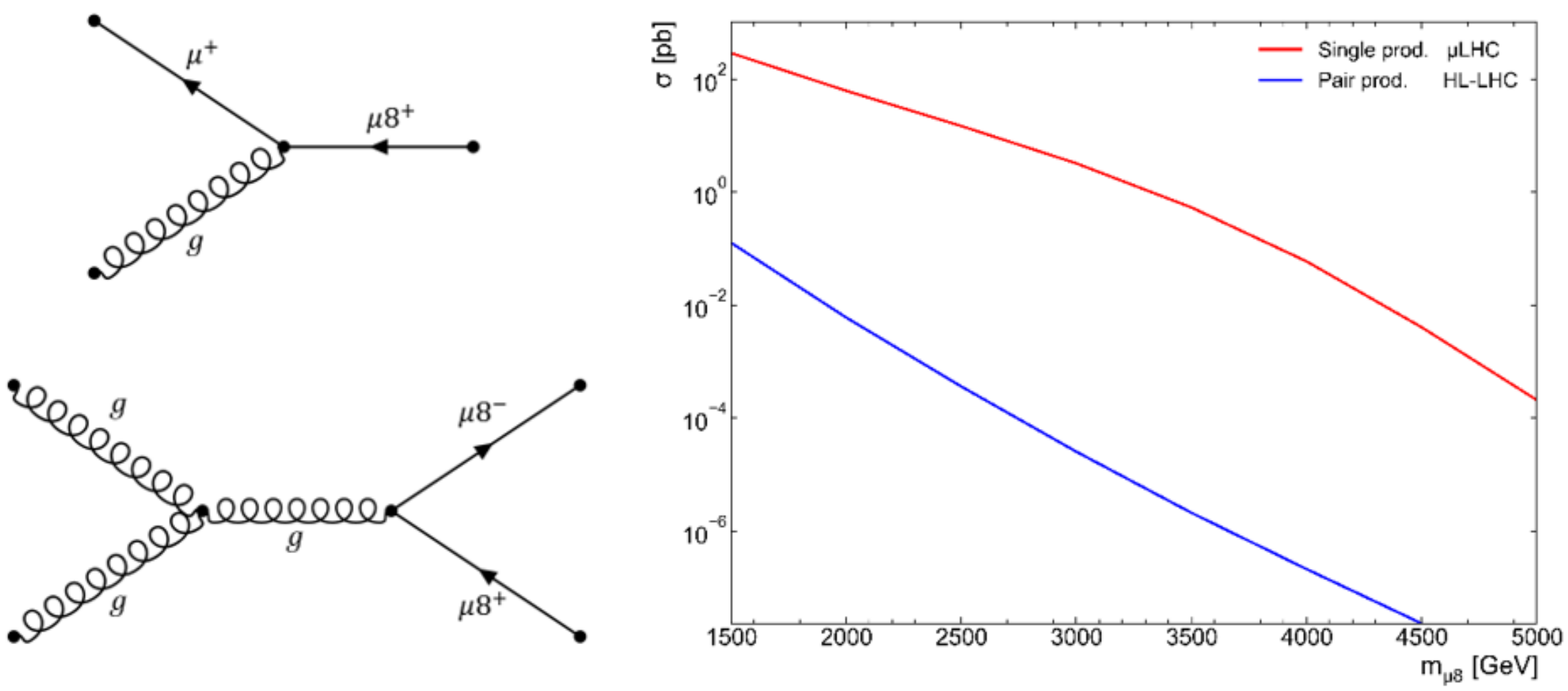

**Figure 13.** $\mu$LHC single production (top left), representative HL-LHC pair production (bottom left) diagrams and cross section versus $\mu 8$ mass plots.

In this study, a hadron level analysis was performed to determine statistical significance values. The $\mu 8$'s formed in the signal events were decayed to a muon and gluon, and the basic backgrounds with the same final states as single production at $\mu$LHC and pair production at HL-LHC were chosen as the backgrounds ($\mu^+ p \rightarrow \mu^+ j$ for single production, $p p \rightarrow \mu^- \mu^+ j j$ for pair production). The total number of generated signal and background events is 100k. Pythia8 [73] was utilized for the hadronization simulation of the generated Monte Carlo (MC) events. The resulting HEPMC files were processed using the jet clustering algorithms with FastJet [74], and the analysis was completed using the MadAnalysis5 [75].

Figure 14 shows the leading jet and leading muon transverse momentum ($p_T$) distributions after pre-selection cuts were applied to events passing through Pythia8 to get events with only 1 jet and 1 lepton in the final state for $\mu$LHC. As evident from the distributions, the $p_T$'s of these particles resulting from the decay of $\mu 8$, which was produced stagnantly in single production, are close to half of the $\mu 8$ masses. As this phenomenon was not detected in the background, a 400 GeV $p_T$ cut was implemented for both the jet and the muon. The pseudo-rapidity cuts $|\eta| < 3.5$ for the jet and $|\eta| < 3$ for the muon were also utilized to ascertain that the particles are in the central region.

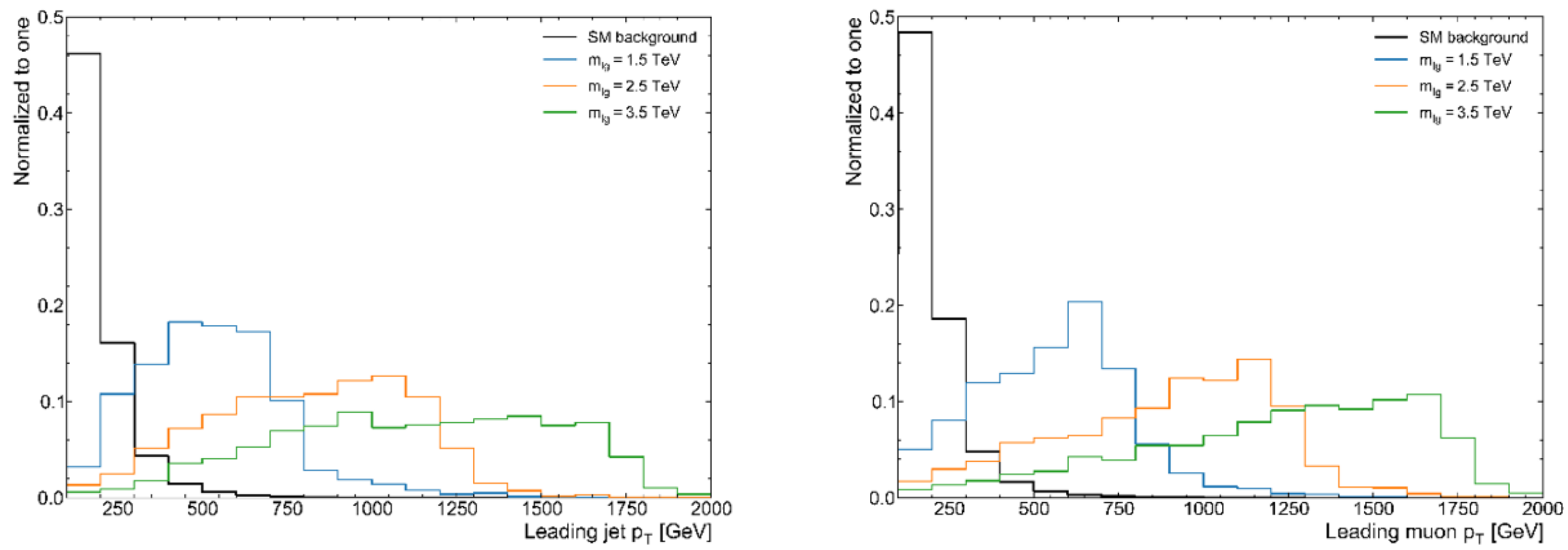

**Figure 14.** $p_T$ distributions of leading jet (left) and leading muon (right) for μLHC signal and background events.

The muon-jet invariant mass distributions obtained after the implementation of the cuts are given in Figure 15. In the determination of these distributions, the integrated luminosity was taken to be 1 $ab^{-1}$.

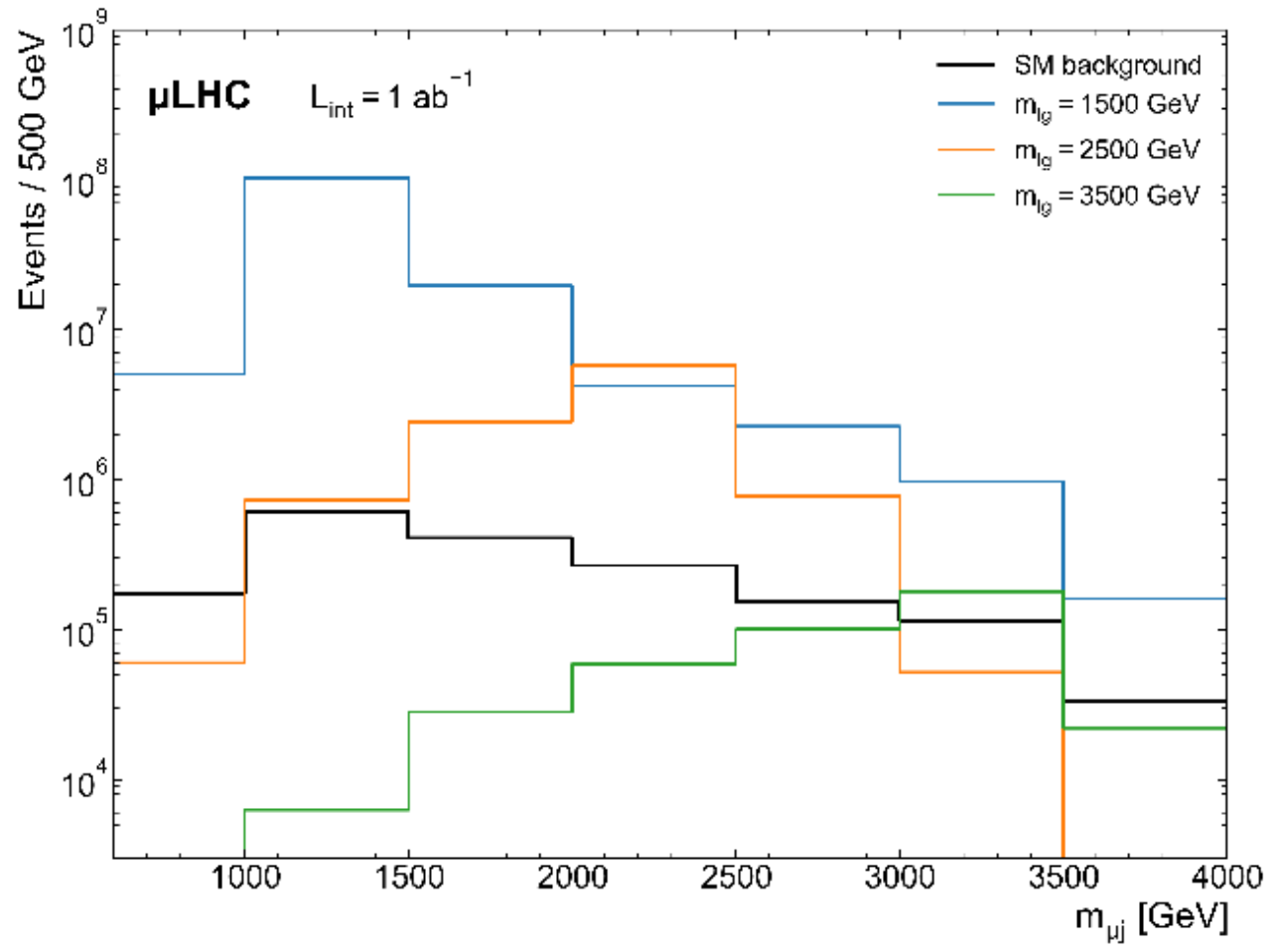

**Figure 15.** The muon-jet invariant mass distributions.

A parallel analysis was conducted for pair production at HL-LHC. Preselection cuts were applied to achieve a final state with 2 muons and 2 jets. Subsequently, the pseudo-rapidity cuts were used at the same level as for $\mu$LHC, while the $p_T$ cuts were made softer: the $p_T$ cuts were set 300 GeV for the leading jet/muon and 250 GeV for the second leading jet/muon.

The statistical significance ($S/\sqrt{B}$, where S and B are the total signal and background event numbers, respectively) graphs of the $\mu 8$ analyses at $\mu$LHC (for various integrated luminosities) and HL-LHC are presented in Figure 16. As can be seen from the figure, $\mu$LHC reaches the same statistical significance values for $\mu 8$ mass as HL-LHC at 100 $nb^{-1}$. This finding indicates that $\mu$LHC has the potential to probe higher masses than HL-LHC in the context of leptogluon research. For example, discovery mass limits ($5\sigma$) at HL-LHC and $\mu$LHC (100 fb$^{-1}$) are approximately 2300 GeV and 4100 GeV, respectively.

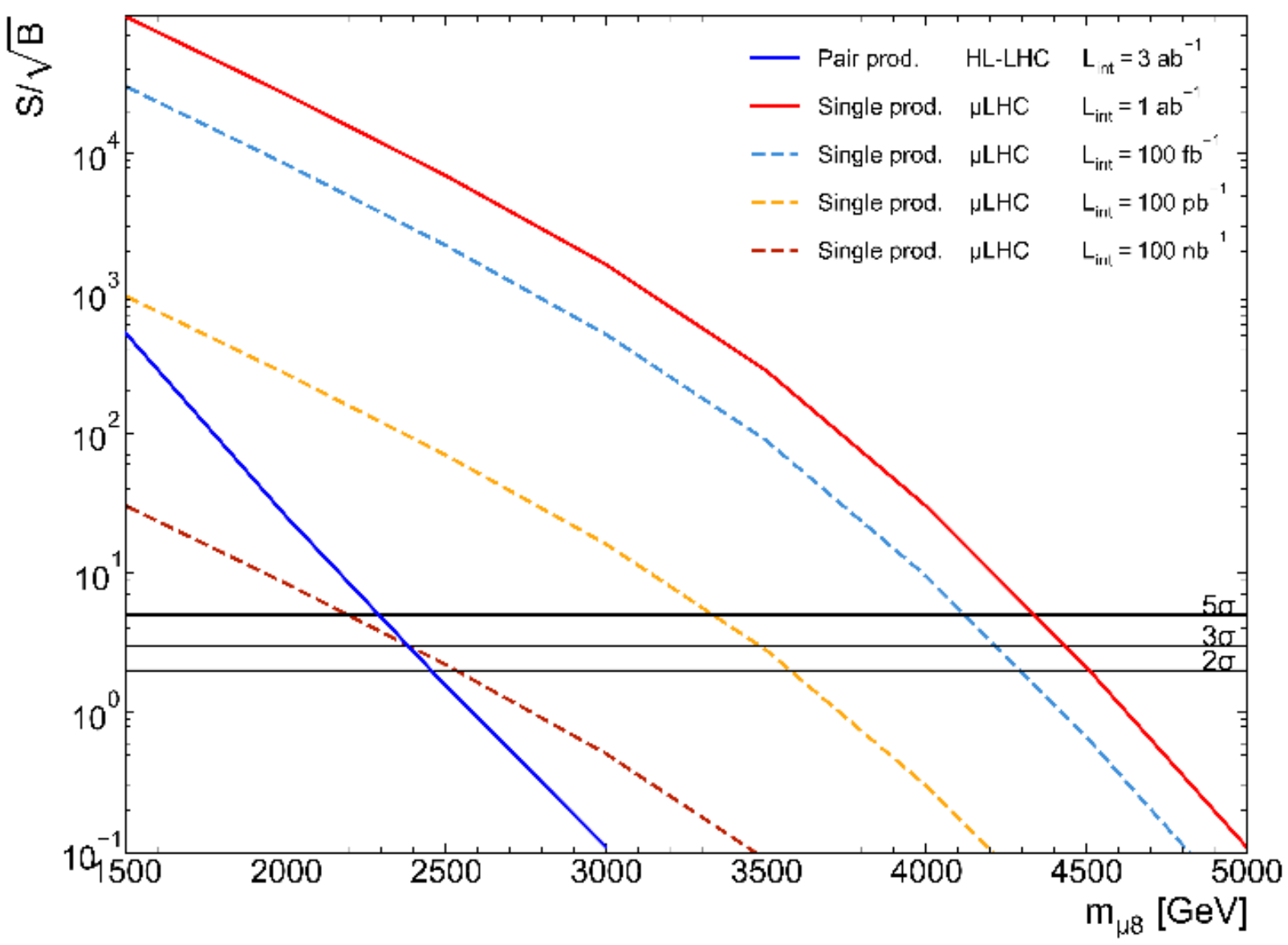

**Figure 16.** Statistical significance graphs.

## 4. Conclusion

This study demonstrates that construction of $\mu$TRISTAN's $\mu^+$-ring tangential to the LHC will provide the opportunity to realize $\mu^+p$ collider with 5.3 TeV center-of-mass energy at luminosity well-exceeding $10^{33}$ cm$^{-2}$s$^{-1}$. Another option would be to use the ERL part of LH*e*C to accelerate the $\mu^+$ beam, in which case it could achieve slightly lower luminosity (but still above $10^{33}$ cm$^{-2}$s$^{-1}$) $\mu^+p$ collisions at the same center-of-mass energy. In fact, further optimization of proton and antimuon beam parameters holds the potential to significantly increase the luminosity value. For instance, the options explored in this study utilized only one-tenth of the available proton bunches. By decreasing the total number of bunches while simultaneously increasing the number of protons per bunch, the luminosity could be enhanced by several times. Such optimization efforts will be the subject of more detailed investigations in future studies.

Obviously, $\mu$LHC will essentially enlarge the physics search potential of the HL-LHC for both the SM and BSM phenomena. As shown in Section 3, $\mu$LHC will make significant contributions to the clarification of QCD basics and Higgs boson properties. Concerning BSM physics, these colliders will provide a huge potential for investigation of muon-related new phenomena, namely excited muon, excited muon neutrino, color-octet muon, leptoquarks, contact interactions, etc. For example, as can be seen from Figure 16, the exploration potential of color-octet muons at the $\mu$LHC is well above that of the HL-LHC. Therefore, systematic studies of accelerator, detector and physics search aspects of the $\mu$LHC is necessary for long-term planning in the field of high energy physics.

Finally, let us emphasize that the antimuon–hadron colliders can be realized before the muon colliders. The reason is that while $\mu^-$ beams with emittance, which is sufficiently small for the construction of muon colliders, have not yet been achieved, there is an established technology to create a low emittance $\mu^+$ beam by using ultra-cold muons. Therefore, $\mu^+p$ colliders might become the second most effective tool, after proton colliders, for exploring the multi-TeV scale at the constituent level.

## Acknowledgments

The authors are grateful to Gokhan Unel for his helpful consultation. We also thank Krzysztof Piotrzkowski for pointing out the inconsistency in antimuon survival rates. As a result, the luminosity value increased by approximately 2 times for the $\mu$TRISTAN-based option and 4 times for the LH$e$C-based option.

## References

[1] R. Hofstadter and R. W. McAllister, Electron Scattering from the Proton, Physical Review **98**, 217 (1955).

[2] R. W. McAllister and R. Hofstadter, Elastic Scattering of 188-Mev Electrons from the Proton and the Alpha Particle, Physical Review **102**, 851 (1956).

[3] E. D. Bloom et al., High-energy inelastic e-p scattering at 6°and 10°, Phys Rev Lett **23**, 930 (1969).

[4] M. Breidenbach, J. I. Friedman, H. W. Kendall, E. D. Bloom, D. H. Coward, H. Destaebler, J. Drees, L. W. Mo, and R. E. Taylor, Observed behavior of highly inelastic electron-proton scattering, Phys Rev Lett **23**, 935 (1969).

[5] J. J. Aubert et al., The ratio of the nucleon structure functions F2N for iron and deuterium, Physics Letters B **123**, 275 (1983).

[6] M. Klein and R. Yoshida, Collider physics at HERA, Prog Part Nucl Phys **61**, 343 (2008).

[7] R. Abdul Khalek et al., Science Requirements and Detector Concepts for the Electron-Ion Collider: EIC Yellow Report, Nucl Phys A **1026**, 122447 (2022).

[8] G. Aad et al., Observation of a new particle in the search for the Standard Model Higgs boson with the ATLAS detector at the LHC, Physics Letters B **716**, 1 (2012).

[9] S. Chatrchyan et al., Observation of a new boson at a mass of 125 GeV with the CMS experiment at the LHC, Physics Letters B **716**, 30 (2012).

[10] G. Apollinari, I. Béjar Alonso, O. Brüning, P. Fessia, M. Lamont, L. Rossi, and L. Tavian, High-luminosity large hadron collider (HL-LHC). Technical design report V.0.1, CERN Yellow Reports: Monographs **4**, 1 (2017).

[11] A. Abada et al., HE-LHC: The High-Energy Large Hadron Collider, The European Physical Journal Special Topics 2019 228:5 **228**, 1109 (2019).

[12] A. Abada et al., FCC-hh: The Hadron Collider, The European Physical Journal Special Topics 2019 228:4 **228**, 755 (2019).

[13] J. Tang, Design Concept for a Future Super Proton-Proton Collider, Front Phys **10**, 828878 (2022).

[14] J. L. Abelleira Fernandez et al., A Large Hadron Electron Collider at CERN Report on the Physics and Design Concepts for Machine and Detector, Journal of Physics G: Nuclear and Particle Physics **39**, 075001 (2012).

[15] P. Agostini et al., The Large Hadron–Electron Collider at the HL-LHC, Journal of Physics G: Nuclear and Particle Physics **48**, 110501 (2021).

[16] V. Shiltsev, An asymmetric μ-p collider as a quark structure microscope: Luminosity consideration, FERMILAB-TM-1969 (1996).

[17] V. Shiltsev, An asymmetric muon-proton collider: luminosity consideration, Conf. Proc. C 970512 420-421 (1997).

[18] I. F. Ginzburg, Physics at future ep, gamma p (linac-ring) and mu p Colliders, Turkish Journal of Physics **22**, 607 (1998).

[19] S. Sultanov, Prospects of the future ep and γp colliders: Luminosity and physics, ICTP Preprint IC/89/409 (1989).

[20] B. Wiik, Recent development in Accelerators, Proceedings, International Europhysics Conference, Marseille, France, July 22-28, 739-758 (1993).

[21] R. Brinkmann, S. Sultansoy, S. Türköz, F. Willeke, O. Yavas, and M. Yılmaz, Linac-ring type colliders: Fourth way to TeV scale, DESY Preprint DESY-97-239; e-Print: Physics/9712023 (1997).

[22] S. Sultansoy, Four ways to TeV scale, Turkish Journal of Physics **22**, 575 (1998).

[23] S. Sultansoy, The post-HERA era: brief review of future lepton-hadron and photon-hadron colliders, DESY Preprint DESY-99-159 (1999); e-Print: Hep-Ph/9911417 [Hep-Ph] (1999).

[24] S. Sultansoy, Linac-ring type colliders: Second way to TeV scale, The European Physical Journal C **33**, 1064 (2004).

[25] S. Sultansoy, A Review of TeV scale lepton-hadron and photon-hadron colliders, Published in: Conf. Proc. C 0505161 (2005) 4329; e-Print: Hep-Ex/0508020 [Hep-Ex] (n.d.).

[26] A. N. Akay, H. Karadeniz, and S. Sultansoy, Review of linac-ring-type collider proposals, International Journal of Modern Physics A **25**, 4589 (2010).

[27] Y. C. Acar, A. N. Akay, S. Beser, A. C. Canbay, H. Karadeniz, U. Kaya, B. B. Oner, and S. Sultansoy, Future circular collider based lepton–hadron and photon–hadron colliders: Luminosity and physics, Nucl Instrum Methods Phys Res A **871**, 47 (2017).

[28] A. C. Canbay, U. Kaya, B. Ketenoglu, B. B. Oner, and S. Sultansoy, SppC Based Energy Frontier Lepton-Proton Colliders: Luminosity and Physics, Advances in High Energy Physics **2017**, 4021493 (2017).

[29] K. Cheung and Z. S. Wang, Physics potential of a muon-proton collider, Physical Review D **103**, 116009 (2021).

[30] D. Acosta and W. Li, A muon–ion collider at BNL: The future QCD frontier and path to a new energy frontier of μ+μ− colliders, Nucl Instrum Methods Phys Res A **1027**, 166334 (2022).

[31] U. Kaya, B. Ketenoglu, S. Sultansoy, and F. Zimmermann, Luminosity and physics considerations on HL-LHC–and HE-LHC–based μp colliders, Europhys Lett **138**, 24002 (2022).

[32] B. Ketenoglu, B. Dagli, A. Ozturk, and S. Sultansoy, Review of muon-proton and muon-nucleus collider proposals, Mod Phys Lett A **37**, (2023).

[33] G. Altarelli, R. Rückl, and B. Melé, Physics of ep collisions in the TeV energy range, Presented at ECFA-CERN Workshop on Feasibility of Hadron Colliders in LEP Tunnel, Lausanne and Geneva, Switzerland; Https://Cds.Cern.Ch/Record/154035/Files/P549.Pdf (1984).

[34] G. Brianti, The large hadron collider in the LEP tunnel, Proceedings of the Workshop on Physics at Future Accelerators, (1987), La Thuile, Italy, CERN-87-07-V-1; Https://Cds.Cern.Ch/Record/194335/Files/6.Pdf (1987).

[35] U. Amaldi, Physics and detectors at the Large Hadron Collider and at the CERN Linear Collider, Proceedings of the Workshop on Physics at Future Accelerators, (1987), La Truille, Italy, Pp.323-352; Https://Cds.Cern.Ch/Record/178303/Files/198706435.Pdf (1987).

[36] S.I.Alekhin et al., Prospects of the Future ep-Colliders, IHEP Preprint 87-48, Serpukhov (1987).

[37] A. Verdier, An ep Insertion for LHC and LEP, Proc. Aachen LHC Workshop, CERN-90-10-B, 820-823 (1990).

[38] E. Keil, LHC ep option, LHC Report 93, CERN (1997).

[39] P. Grosse-Wiesmann, Colliding a linear electron beam with a storage ring beam, Nucl Instrum Methods Phys Res A **274**, 21 (1989).

[40] M. Tigner, B. Wiik, and F. Willeke, An Electron - proton collider in the TeV range, Conf.Proc.C **910506**, 2910 (1991).

[41] Z. Z. Aydin, A. K. Çiftçi, and S. Sultansoy, Linac-ring type ep machines and γp colliders based on them, Nucl Instrum Methods Phys Res A **351**, 261 (1994).

[42] O. Yavas, A. Çiftçi, and S. Sultansoy, Linac* LHC based ep, gamma-p, eA, gamma-A and FEL gamma-A Colliders, Proceedings of the 7th European Conference, EPAC 2000, Vienna, Austria, June 26-30, 2000, 391-396; e-Print: Hep-Ex/0004016 [Hep-Ex] (2000).

[43] D. Schulte and F. Zimmermann, QCD Explorer based on LHC and CLIC-1, 9th European Particle Accelerator Conference (EPAC 2004), Proceedings of EPAC 2004, Lucerne, Switzerland, p. 632-634 (2004).

[44] *A Large Hadron Electron Collider at CERN*, https://lhec.web.cern.ch/.

[45] U. Kaya, B. Ketenoğlu, and S. Sultansoy, The LHeC project: e-Ring revisited, Süleyman Demirel University Faculty of Arts and Science Journal of Science **13**, 173 (2018).
[46] A. N. Akay, B. Dağli, B. Ketenoğlu, A. Öztürk, and S. Sultansoy, Alternative scenarios for the LHC based electron-proton collider, Turkish Journal of Physics **48**, 142 (2024).
[47] J. P. Delahaye, M. Diemoz, K. Long, B. Mansoulié, N. Pastrone, L. Rivkin, D. Schulte, A. Skrinsky, and A. Wulzer, Muon Colliders, ArXiv:1901.06150 [Physics.Acc-Ph] (2019).
[48] C. Accettura et al., Towards a muon collider, The European Physical Journal C 2023 83:9 **83**, 1 (2023).
[49] C. Accettura et al., The Muon Collider, E-Print: ArXiv:2504.21417 (2025).
[50] Y. Kondo, R. Kitamura, S. Li, S. Bae, H. Choi, S. Choi, B. Kim, H. S. Ko, and G. P. Razuvaev, Re-acceleration of ultra cold muon in JPARC muon facility, 10.18429/JACoW-IPAC2018-FRXGBF1 (2018).
[51] Y. Hamada, R. Kitano, R. Matsudo, H. Takaura, and M. Yoshida, μTRISTAN, Progress of Theoretical and Experimental Physics **2022**, 53 (2022).
[52] D. Akturk, B. Dagli, and S. Sultansoy, Muon ring and FCC-ee/CEPC based antimuon-electron colliders, Europhys Lett **147**, 24001 (2024).
[53] D. Akturk, B. Dagli, B. Ketenoglu, A. Ozturk, and S. Sultansoy, μTRISTAN- and LHC-/Tevatron-/FCC-/SppC-based antimuon–hadron colliders, The European Physical Journal Plus 2025 140:4 **140**, 1 (2025).
[54] *Overview | J-PARC Muon g-2/EDM Experiment*, https://g-2.kek.jp/overview/.
[55] K. André, B. Holzer, L. Forthomme, and K. Piotrzkowski, An electron-hadron collider at the high-luminosity LHC, ArXiv:2503.20475v1 [Hep-Ex] (2025).
[56] Black, K. M., Jindariani, S., Li, D., Maltoni, F., Meade, P., Stratakis, D., Acosta, D., Agarwal, R., Agashe, K., Aimè, C., et al. (2024). Muon collider forum report. *Journal of Instrumentation*, *19*(02), T02015. https://doi.org/10.1088/1748-0221/19/02/T02015
[57] D. Acosta, E. Barberis, N. Hurley, W. Li, O. Miguel Colin, Y. Wang, D. Wood, and X. Zuo, The potential of a TeV-scale muon-ion collider, Journal of Instrumentation **18**, P09025 (2023).
[58] Taylor, R., Chancé, A., Giove, D. A., Milas, N., Losito, R., Lucchesi, D., Rogers, C., Rossi, L., Schulte, D., Accettura, C., et al. (2025). MuCol Milestone Report No. 7: Consolidated parameters. *arXiv preprint arXiv:2510.27437*. https://doi.org/10.48550/arXiv.2510.27437
[59] Topp-Mugglestone, M. E. (2024). *Scaling fixed field accelerators: theory and modelling of horizontal-and vertical-excursion accelerators* [PhD thesis, University of Oxford]. Oxford University Research Archive.
[60] B. Dagli, S. Sultansoy, B. Ketenoglu, and B. Oner, *Beam-Beam Simulations for Lepton-Hadron Colliders: ALOHEP Software*, in *12th Int. Particle Accelerator Conf.* (2021).
[61] *GitHub - Yefetu/ALOHEP: ALOHEP(A Luminosity Optimizer for High Energy Physics)*, https://github.com/yefetu/ALOHEP.
[62] *High Energy Physics*, http://yef.etu.edu.tr/ALOHEP2_eng.html.
[63] D. Acosta, E. Barberis, M. N. Hurley, M. W. Li, M. O. M. Colin, M. M. Munyi, Y. Wang, M. D. Wood, M. K. Yang, and X. Zuo, Physics Potential, Accelerator Options, and Experimental Challenges of a TeV-Scale Muon-Ion Collider, PoS(ICHEP2024)817 (2024).
[64] H. Davoudiasl, H. Liu, R. Marcarelli, Y. Soreq, and S. Trifinopoulos, New physics at the Muon (Synchrotron) Ion Collider: MuSIC for several scales, Journal of High Energy Physics 2025 2025:3 **2025**, 1 (2025).
[65] S. Navas et al., Review of Particle Physics, Physical Review D **110**, 030001 (2024).
[66] T. Von Witzleben, K. D. J. André, R. De Maria, B. Holzer, M. Klein, J. Pretz, and M. Smith, JACOW: Beam dynamics for concurrent operation of the LHeC and the HL-LHC, 14th International Particle Accelerator Conference (2023).
[67] I. D'Souza and C. Kalman, *Preons: Models of Leptons, Quarks and Gauge Bosons as Composite Objects* (World Scientific Publishing, Hong Kong. https://doi.org/10.1142/9789814354769, 1992).
[68] D. Gonçalves-Netto, D. López-Val, K. Mawatari, I. Wigmore, and T. Plehn, Looking for leptogluons, Physical Review D - Particles, Fields, Gravitation and Cosmology **87**, (2013).
[69] Y. C. Acar, U. Kaya, and B. B. Oner, Resonant production of color octet muons at Future Circular Collider-based muon-proton colliders*, Chinese Physics C **42**, 083108 (2018).
[70] J. Alwall, M. Herquet, F. Maltoni, O. Mattelaerc, and T. Stelzerd, MadGraph 5: Going beyond, Journal of High Energy Physics **2011**, 1 (2011).
[71] *A.C. Canbay and U. Kaya, github − acanbay/leptogluon 2ndgen ufo (2025).* https://github.com/acanbay/Leptogluon_2ndgen_UFO.
[72] N. D. Christensen and C. Duhr, FeynRules – Feynman rules made easy, Comput Phys Commun **180**, 1614 (2009).

[73] T. Sjöstrand, S. Mrenna, and P. Skands, A brief introduction to PYTHIA 8.1, Comput Phys Commun **178**, 852 (2008).
[74] M. Cacciari, G. P. Salam, and G. Soyez, FastJet user manual, The European Physical Journal C 2012 72:3 **72**, 1 (2012).
[75] E. Conte, B. Fuks, and G. Serret, MadAnalysis 5, a user-friendly framework for collider phenomenology, Comput Phys Commun **184**, 222 (2013).